\documentclass[11pt,a4paper]{article}

\usepackage[margin=1in]{geometry}
\usepackage{authblk}
\usepackage{graphicx}
\usepackage{multirow}
\usepackage{amsmath,amssymb,amsfonts}
\usepackage{booktabs}
\usepackage{siunitx}
\usepackage[T1]{fontenc}
\usepackage{hyperref}
\usepackage{orcidlink}
\usepackage[super,sort&compress,numbers]{natbib}
\usepackage{xspace}
\xspaceaddexceptions{\% \textdegree}
\usepackage{caption}
\newcommand{\numSubjects}{965\xspace}
\newcommand{\nAgeGroupYoung}{295\xspace}
\newcommand{\nAgeGroupMid}{259\xspace}
\newcommand{\nAgeGroupOlder}{257\xspace}
\newcommand{\nAgeGroupSenior}{154\xspace}
\newcommand{\nFemale}{483\xspace}
\newcommand{\nMale}{482\xspace}
\newcommand{\numSkinConditions}{6\xspace}
\newcommand{\nSkinNeutral}{234\xspace}
\newcommand{\nSkinDry}{279\xspace}
\newcommand{\nSkinSeverelyDry}{3\xspace}
\newcommand{\nSkinComboDry}{139\xspace}
\newcommand{\nSkinComboOily}{142\xspace}
\newcommand{\nSkinOily}{168\xspace}
\newcommand{\numFacialRegionsAnalyzed}{5\xspace}
\newcommand{\numDslrAngles}{7\xspace}
\newcommand{\totalSampleRecords}{43{,}425\xspace}
\newcommand{\analyzedSampleRecords}{43{,}424\xspace}
\newcommand{\nPublicReleaseImages}{13{,}936\xspace}
\newcommand{\nDslrImages}{7{,}504\xspace}
\newcommand{\nTabletImages}{3{,}216\xspace}
\newcommand{\nSmartphoneImages}{3{,}216\xspace}
\newcommand{\totalCaptureSessions}{1{,}072\xspace}
\newcommand{\completeSessions}{965\xspace}
\newcommand{\droppedSessions}{107\xspace}
\newcommand{\nDslrSamplesAllAngles}{33{,}774\xspace}
\newcommand{\nFrontPerDevice}{4{,}825\xspace}
\newcommand{\nCrossDeviceTotal}{14{,}475\xspace}
\newcommand{\nAnovaPairs}{9{,}650\xspace}
\newcommand{\numConsumerAngles}{3\xspace}
\newcommand{\totalAnatomicalRegions}{8\xspace}
\newcommand{\numDevices}{3\xspace}
\newcommand{\viewAngleSmall}{15\xspace}
\newcommand{\viewAngleLarge}{30\xspace}

\newcommand{\meanLstarRef}{65\xspace}

\newcommand{\meanLstarDslr}{64.9\xspace}
\newcommand{\meanLstarTablet}{53.2\xspace}
\newcommand{\meanLstarSmartphone}{60.6\xspace}
\newcommand{\meanAstarDslr}{17.9\xspace}
\newcommand{\meanAstarTablet}{10.5\xspace}
\newcommand{\meanAstarSmartphone}{10.1\xspace}
\newcommand{\meanBstarDslr}{29.1\xspace}
\newcommand{\meanBstarTablet}{15.0\xspace}
\newcommand{\meanBstarSmartphone}{14.1\xspace}

\newcommand{\deltaEMeanTabletDslr}{13.27\xspace}
\newcommand{\deltaEStdTabletDslr}{4.28\xspace}
\newcommand{\deltaEMedianTabletDslr}{12.55\xspace}
\newcommand{\deltaEPctileTabletDslr}{21.30\xspace}
\newcommand{\deltaEMeanSmartphoneDslr}{10.63\xspace}
\newcommand{\deltaEStdSmartphoneDslr}{2.85\xspace}
\newcommand{\deltaEMedianSmartphoneDslr}{10.30\xspace}
\newcommand{\deltaEPctileSmartphoneDslr}{15.73\xspace}
\newcommand{\deltaEMeanTabletSmartphone}{7.39\xspace}
\newcommand{\deltaEStdTabletSmartphone}{3.81\xspace}
\newcommand{\deltaEMedianTabletSmartphone}{7.02\xspace}
\newcommand{\deltaEPctileTabletSmartphone}{14.25\xspace}
\newcommand{\deltaEMinSmartphoneDslr}{3.45\xspace}
\newcommand{\deltaEFoldExceedSmartphone}{5.3\xspace}
\newcommand{\deltaEFoldExceedTablet}{6.6\xspace}

\newcommand{\ccmTabletImprovementPct}{73.9\xspace}
\newcommand{\ccmSmartphoneImprovementPct}{61.1\xspace}
\newcommand{\ccmReductionLow}{61\xspace}
\newcommand{\ccmReductionHigh}{74\xspace}
\newcommand{\deltaETabletPostCcmMean}{3.46\xspace}
\newcommand{\deltaETabletPostCcmStd}{2.36\xspace}
\newcommand{\deltaETabletPostCcmMedian}{2.90\xspace}
\newcommand{\deltaESmartphonePostCcmMean}{4.14\xspace}
\newcommand{\deltaESmartphonePostCcmStd}{2.56\xspace}

\newcommand{\iccMelaninGlobal}{0.797\xspace}
\newcommand{\iccItaGlobal}{0.776\xspace}
\newcommand{\iccLstarGlobal}{0.783\xspace}
\newcommand{\iccAstarGlobal}{0.725\xspace}
\newcommand{\iccBstarGlobal}{0.713\xspace}
\newcommand{\iccMelaninGlobalCI}{[0.79, 0.81]\xspace}
\newcommand{\iccItaGlobalCI}{[0.77, 0.79]\xspace}
\newcommand{\iccLstarGlobalCI}{[0.77, 0.79]\xspace}
\newcommand{\iccAstarGlobalCI}{[0.71, 0.74]\xspace}
\newcommand{\iccBstarGlobalCI}{[0.70, 0.72]\xspace}
\newcommand{\iccMelaninAbstract}{0.80\xspace}
\newcommand{\iccItaAbstract}{0.78\xspace}

\newcommand{\loaSdMultiplier}{1.96\xspace}

\newcommand{\anovaFRegion}{622.79\xspace}
\newcommand{\anovaPRegion}{< 10^{-300}\xspace}
\newcommand{\anovaEtaRegion}{0.1772\xspace}
\newcommand{\anovaFDevice}{1638.43\xspace}
\newcommand{\anovaPDevice}{< 10^{-300}\xspace}
\newcommand{\anovaEtaDevice}{0.1165\xspace}
\newcommand{\anovaFSex}{273.76\xspace}
\newcommand{\anovaPSex}{1.18 \times 10^{-60}\xspace}
\newcommand{\anovaEtaSex}{0.0195\xspace}
\newcommand{\anovaFSkin}{1.77\xspace}
\newcommand{\anovaPSkin}{0.115\xspace}
\newcommand{\anovaEtaSkin}{0.0006\xspace}
\newcommand{\anovaFAge}{4.07\xspace}
\newcommand{\anovaPAge}{0.007\xspace}
\newcommand{\anovaEtaAge}{0.0009\xspace}
\newcommand{\anovaEtaRegionAbstract}{0.18\xspace}
\newcommand{\anovaEtaDeviceAbstract}{0.12\xspace}
\newcommand{\anovaEtaRegionPct}{17.7\xspace}
\newcommand{\anovaRegionDeviceRatio}{1.5\xspace}

\newcommand{\misclassTabletGlobal}{39.2\xspace}
\newcommand{\misclassSmartphoneGlobal}{45.0\xspace}
\newcommand{\misclassTabletRegional}{19.9\xspace}
\newcommand{\misclassSmartphoneRegional}{27.5\xspace}
\newcommand{\misclassForeheadTabletGlobal}{42.1\xspace}
\newcommand{\misclassLcheekTabletGlobal}{37.2\xspace}
\newcommand{\misclassRcheekTabletGlobal}{39.2\xspace}
\newcommand{\misclassChinTabletGlobal}{35.5\xspace}
\newcommand{\misclassGlabellaTabletGlobal}{42.0\xspace}
\newcommand{\misclassForeheadTabletRegional}{23.2\xspace}
\newcommand{\misclassLcheekTabletRegional}{22.5\xspace}
\newcommand{\misclassRcheekTabletRegional}{23.0\xspace}
\newcommand{\misclassChinTabletRegional}{14.7\xspace}
\newcommand{\misclassGlabellaTabletRegional}{16.0\xspace}
\newcommand{\misclassForeheadSmartphoneGlobal}{45.9\xspace}
\newcommand{\misclassLcheekSmartphoneGlobal}{41.8\xspace}
\newcommand{\misclassRcheekSmartphoneGlobal}{43.0\xspace}
\newcommand{\misclassChinSmartphoneGlobal}{46.0\xspace}
\newcommand{\misclassGlabellaSmartphoneGlobal}{48.5\xspace}
\newcommand{\misclassForeheadSmartphoneRegional}{40.1\xspace}
\newcommand{\misclassLcheekSmartphoneRegional}{30.5\xspace}
\newcommand{\misclassRcheekSmartphoneRegional}{33.0\xspace}
\newcommand{\misclassChinSmartphoneRegional}{14.7\xspace}
\newcommand{\misclassGlabellaSmartphoneRegional}{19.1\xspace}
\newcommand{\misclassAdjacentPct}{98.3\xspace}
\newcommand{\misclassPerRegionLowGlobal}{35\xspace}
\newcommand{\misclassPerRegionHighGlobal}{49\xspace}

\newcommand{\iccMelaninRegional}{0.946\xspace}
\newcommand{\iccItaRegional}{0.926\xspace}
\newcommand{\iccItaRegionalAbstract}{0.93\xspace}
\newcommand{\iccMelaninRegionalAbstract}{0.95\xspace}
\newcommand{\deltaETabletGlobalCcmRegionalMean}{3.46\xspace}
\newcommand{\deltaETabletRegionalCcmRegionalMean}{2.23\xspace}
\newcommand{\deltaESmartphoneGlobalCcmRegionalMean}{4.14\xspace}
\newcommand{\deltaESmartphoneRegionalCcmRegionalMean}{2.68\xspace}
\newcommand{\regionalCcmReductionPctTablet}{36\xspace}
\newcommand{\regionalCcmReductionPctSmartphone}{35\xspace}
\newcommand{\deltaEForeheadTabletGlobal}{2.95\xspace}
\newcommand{\deltaEForeheadTabletRegional}{1.94\xspace}
\newcommand{\deltaEForeheadTabletRegionalReductionPct}{34.2\xspace}
\newcommand{\deltaELcheekTabletGlobal}{2.92\xspace}
\newcommand{\deltaELcheekTabletRegional}{1.90\xspace}
\newcommand{\deltaELcheekTabletRegionalReductionPct}{34.9\xspace}
\newcommand{\deltaERcheekTabletGlobal}{2.90\xspace}
\newcommand{\deltaERcheekTabletRegional}{1.99\xspace}
\newcommand{\deltaERcheekTabletRegionalReductionPct}{31.4\xspace}
\newcommand{\deltaEChinTabletGlobal}{4.64\xspace}
\newcommand{\deltaEChinTabletRegional}{3.45\xspace}
\newcommand{\deltaEChinTabletRegionalReductionPct}{25.6\xspace}
\newcommand{\deltaEGlabellaTabletGlobal}{3.91\xspace}
\newcommand{\deltaEGlabellaTabletRegional}{1.89\xspace}
\newcommand{\deltaEGlabellaTabletRegionalReductionPct}{51.7\xspace}
\newcommand{\deltaEForeheadSmartphoneGlobal}{4.19\xspace}
\newcommand{\deltaEForeheadSmartphoneRegional}{2.53\xspace}
\newcommand{\deltaEForeheadSmartphoneRegionalReductionPct}{39.6\xspace}
\newcommand{\deltaELcheekSmartphoneGlobal}{3.37\xspace}
\newcommand{\deltaELcheekSmartphoneRegional}{2.42\xspace}
\newcommand{\deltaELcheekSmartphoneRegionalReductionPct}{28.2\xspace}
\newcommand{\deltaERcheekSmartphoneGlobal}{3.24\xspace}
\newcommand{\deltaERcheekSmartphoneRegional}{2.54\xspace}
\newcommand{\deltaERcheekSmartphoneRegionalReductionPct}{21.4\xspace}
\newcommand{\deltaEChinSmartphoneGlobal}{5.40\xspace}
\newcommand{\deltaEChinSmartphoneRegional}{3.58\xspace}
\newcommand{\deltaEChinSmartphoneRegionalReductionPct}{33.7\xspace}
\newcommand{\deltaEGlabellaSmartphoneGlobal}{4.48\xspace}
\newcommand{\deltaEGlabellaSmartphoneRegional}{2.33\xspace}
\newcommand{\deltaEGlabellaSmartphoneRegionalReductionPct}{47.9\xspace}
\newcommand{\deltaEOtherRegionsTabletLow}{1.89\xspace}
\newcommand{\deltaEOtherRegionsTabletHigh}{1.99\xspace}
\newcommand{\deltaEOtherRegionsSmartphoneLow}{2.33\xspace}
\newcommand{\deltaEOtherRegionsSmartphoneHigh}{2.54\xspace}

\newcommand{\iccDslrCeilingLow}{0.86\xspace}
\newcommand{\iccDslrCeilingHigh}{0.94\xspace}
\newcommand{\meanDslrAngularDeltaE}{2.1\xspace}

\newcommand{\meanDeltaERoiChin}{9.40\xspace}
\newcommand{\meanDeltaERoiNonChin}{0.91\xspace}
\newcommand{\iccRobustRoiMiSimple}{0.78\xspace}
\newcommand{\iccRobustRoiMiRobust}{0.52\xspace}
\newcommand{\iccRobustRoiItaSimple}{0.72\xspace}
\newcommand{\iccRobustRoiItaRobust}{0.48\xspace}
\newcommand{\varBetweenMiReductionPct}{66\xspace}
\newcommand{\varBetweenItaReductionPct}{65\xspace}

\newcommand{\deltaEThreshold}{2\xspace}
\newcommand{\blandAltmanMinSamples}{100\xspace}
\newcommand{\crossValidationFolds}{5\xspace}
\newcommand{\alphaSig}{0.05\xspace}
\newcommand{\alphaBonferroni}{0.01\xspace}
\newcommand{\numAnovaTests}{5\xspace}
\newcommand{\cieObserverDeg}{2\xspace}
\newcommand{\srgbKnee}{0.04045\xspace}
\newcommand{\srgbGamma}{2.4\xspace}

\newcommand{\consumerVsCrossConsumerRatioLow}{1.4\xspace}
\newcommand{\consumerVsCrossConsumerRatioHigh}{1.8\xspace}

\title{Region-Specific Calibration Achieves Excellent Inter-Device Reliability for Smartphone Dermatology: A Multi-Device Benchmark on Korean Facial Skin}

\author[1]{Sungwoo Kang\,\orcidlink{0009-0004-0037-7593}\thanks{Corresponding author: \href{mailto:krml919@korea.ac.kr}{krml919@korea.ac.kr}}}
\author[2]{Jong-Kook Kim\,\orcidlink{0000-0003-1828-7807}\thanks{\href{mailto:jongkook@korea.ac.kr}{jongkook@korea.ac.kr}}}
\affil[1]{Department of Electrical and Computer Engineering, Korea University, Seoul, Republic of Korea}
\affil[2]{School of Electrical Engineering, Korea University, Seoul, Republic of Korea}
\date{}

\begin{document}

\maketitle

\begin{abstract}
\noindent
\textbf{Background:} Smartphone-based dermatology requires inter-device colorimetric reliability that holds across calibration regimes, yet quantitative multi-device benchmarks remain scarce.
\textbf{Materials and Methods:} We analyzed matched facial images from \numSubjects Korean subjects captured by a digital single-lens reflex (DSLR) camera, a consumer tablet, and a consumer smartphone, and evaluated two calibration methods against the DSLR reference. The methods are standard global linear Color Correction Matrix (CCM) normalization and region-specific CCM trained per anatomical region, both applied in Commission Internationale de l'\'Eclairage L*a*b* (CIELAB) space.
\textbf{Results:} Linear CCM reduced inter-device color differences by \ccmReductionLow--\ccmReductionHigh\% and placed both Melanin Index (intraclass correlation coefficient [ICC] = \iccMelaninAbstract) and Individual Typology Angle (ITA, ICC = \iccItaAbstract) in the good reliability band. Region-specific CCM raised both indices into the excellent reliability band (MI ICC = \iccMelaninRegionalAbstract, ITA ICC = \iccItaRegionalAbstract), with anatomical region exceeding the source device as the largest pre-calibration variance contributor (analysis-of-variance $\eta^2 = \anovaEtaRegionAbstract$ versus $\anovaEtaDeviceAbstract$).
\textbf{Conclusion:} Consumer-device skin colorimetry therefore achieves clinically useful inter-device reliability using standard calibration, with region-aware calibration the largest remaining source of improvement.
\end{abstract}


\section{Introduction}\label{sec:intro}

Recent Smartphone-based dermatological applications enable remote skin health monitoring at population scale.\cite{Esteva2017} Deep learning systems have demonstrated expert-level performance in melanoma classification,\cite{Marchetti2018,Haenssle2018,Brinker2019} raising the prospect that consumer cameras could serve as quantitative skin-analysis platforms whose downstream clinical indices reproduce across hardware. That prospect rests on inter-device colorimetric reliability, the property that the same skin region yields the same numerical index across cameras after calibration. Whether CCM-calibrated consumer hardware can deliver this reliability at the level of clinical indices, across a fleet of devices sharing a common reference, remains unresolved.

CCM normalization transforms device red--green--blue (RGB) values toward a reference standard in the Commission Internationale de l'\'Eclairage L*a*b* (CIELAB) color space,\cite{Hong2001,Karaimer2018,Luo2001} a perceptually uniform space whose three coordinates capture lightness ($L^*$), red--green chromaticity ($a^*$), and yellow--blue chromaticity ($b^*$). The perceptual color error $\Delta E$ computed using the CIEDE2000 formula\cite{Luo2001} quantifies the difference between corrected and reference colors, with $\Delta E < \deltaEThreshold$ taken as the standard for clinical acceptability, sitting just below the approximate just-noticeable difference ($\Delta E \approx 2.3$) in human color perception established by Mahy et al.\cite{Mahy1994} What CCM normalization achieves on $\Delta E$ is well documented at the level of individual devices.\cite{Hong2001,Karaimer2018} What it achieves at the level of clinical indices computed across a fleet of consumer devices, against a common reference, is less so.

Three properties of the imaging chain make this an empirical question. First, skin color arises from the interaction of light with melanin in the epidermis and hemoglobin in the dermal vasculature, producing a continuous spectral signal that RGB sensors compress into three channels by integrating across wavelength.\cite{Anderson1981,Zonios2001} Second, facial skin varies in epidermal thickness, sebaceous gland density, and subsurface vasculature across anatomical regions,\cite{Chopra2015,Song2019} introducing spatial heterogeneity that a single global correction cannot capture. Third, different cameras distort the already-lossy signal through variations in sensor spectral sensitivities and image-signal-processing pipelines.\cite{Karaimer2016,Hong2001,Wang2018} A multi-device benchmark therefore needs to address all three jointly.

Two clinical indices serve as outcome measures in this study. The Individual Typology Angle (ITA) is the angular skin-type correlate adopted in dermatology,\cite{Fitzpatrick1988,Chardon1991} computed as the arctangent of the ratio of lightness deviation ($L^* - 50$) to yellow--blue chromaticity ($b^*$). The Melanin Index (MI) is the standard CIELAB surrogate for skin pigmentation, a logarithmic function of $L^*$ adapted from Takiwaki's reflectance-based formulation\cite{Takiwaki1998} and its digital-imaging extension by Yamamoto et al.\cite{Yamamoto2008} ITA and MI thus differ in their CIELAB dependencies (ITA on both $L^*$ and $b^*$, MI on $L^*$ only) and in their algebraic forms (an arctangent ratio versus a logarithmic transform). Quantifying inter-device agreement on both indices side by side using the same calibration is the most direct way to test how reliably each transfers across consumer hardware.  Prior work has not tested consumer cameras using canonical-sRGB CCM calibration. The inter-device reliability of ITA on standard consumer cameras remains unmeasured, and the dominant residual variance source is unidentified.

We address both the unmeasured inter-device reliability of ITA and the unidentified dominant variance source through a multi-device intraclass correlation coefficient (ICC) benchmark of \numSubjects Korean subjects (mean reference $L^* \approx \meanLstarRef$ using canonical IEC 61966-2-1 sRGB), comparing two calibration regimes on the same matched samples. The regimes are standard global CCM and region-specific CCM trained separately per anatomical region. The benchmark yields two findings. First, standard global CCM using canonical sRGB delivers good inter-device reliability for both indices, with MI ICC = \iccMelaninAbstract and ITA ICC = \iccItaAbstract. Second, anatomical region edges the source device as the largest pre-calibration variance contributor. Region-specific CCM raises both clinical indices into the excellent band (MI \iccMelaninRegionalAbstract, ITA \iccItaRegionalAbstract) and reduces ITA-based skin-type misclassification, identifying region-aware calibration as the largest remaining source of improvement.

The remainder of the paper details the dataset used and analysis methods, presents the multi-device benchmark and the region-specific calibration result, and discusses the deployment implications for smartphone dermatology.

\section{Methods}\label{sec:methods}

\subsection{Dataset}\label{sec:dataset}

This study utilized the Korean Skin Condition Measurement dataset from AI Hub (aihub.or.kr), a publicly available resource for dermatological research.\cite{AIHub2023} The dataset comprises standardized facial images captured using the AI Hub acquisition protocol (described below) from \numSubjects subjects distributed across four age groups, with sample sizes of n=\nAgeGroupYoung ($\leq$~30 years), n=\nAgeGroupMid (31--45 years), n=\nAgeGroupOlder (46--60 years), and n=\nAgeGroupSenior ($>$60 years). Ages are recorded as integers in the source labels; for stratified analyses we re-aggregated the AI Hub decade groupings (10s, 20s, 30s, 40s, 50s, 60+) into the four bins above using right-closed intervals, so subjects aged exactly 30, 45, and 60 fall in the $\leq$30, 31--45, and 46--60 bins, respectively. The cohort was sex-balanced (\nFemale female, \nMale male). Each subject is additionally annotated with a \numSkinConditions-level skin condition class describing sebum/moisture phenotype, with codes (per the AI Hub dataset specification\cite{AIHub2023}) 0 = neutral, 1 = dry, 2 = severely dry, 3 = combination dry, 4 = combination oily, and 5 = oily, distributed across the cohort as n = \nSkinNeutral, \nSkinDry, \nSkinSeverelyDry, \nSkinComboDry, \nSkinComboOily, and \nSkinOily respectively. The cohort's measured reference colorimetry using canonical IEC 61966-2-1 sRGB conversion (mean $L^* \approx \meanLstarRef$) characterizes the population's skin lightness range.\cite{Choi2017}

Three device categories were employed. The first was a professional-grade DSLR camera (reference) capturing 7 viewing angles (front, tilt-up, tilt-down, left 15\textdegree, left 30\textdegree, right 15\textdegree, right 30\textdegree; ``tilt-up'' and ``tilt-down'' denote pure pitch-axis rotations relative to frontal). The second was a consumer-grade tablet camera capturing 3 viewing angles (front, left, right). The third was a consumer-grade smartphone camera capturing 3 viewing angles (front, left, right). Per the AI Hub dataset documentation,\cite{AIHub2023} the DSLR captures were acquired in a measurement room equipped with lighting equipment (specific illuminant, color temperature, and illuminance levels are not disclosed by the dataset providers), while the tablet and smartphone captures were acquired under generic ceiling lighting in a waiting room. The lighting conditions for the consumer devices are therefore representative of routine indoor environments rather than a calibrated optical rig. The dataset does not disclose specific device makes or models, and devices are identified by category only.

\numFacialRegionsAnalyzed of the dataset's \totalAnatomicalRegions annotated facial regions were analyzed per image: forehead, left cheek, right cheek, chin, and glabella. The remaining three regions---left perocular, right perocular, and lip---were excluded a priori because eyelashes, eyebrow hair, and lip vermilion contaminate skin-colour measurements and are not representative of cutaneous chromophore content. After restricting to these \numFacialRegionsAnalyzed facial regions and the 9 capture configurations used in this study (7 DSLR angles plus the front-facing capture from each consumer device), the full grid contained \totalSampleRecords sample-level records (\numSubjects subjects $\times$ \numFacialRegionsAnalyzed regions $\times$ 9 captures). The public release additionally contains tablet and smartphone left- and right-oblique captures, but their angle codes (a generic ``L'' and ``R'') do not have a documented yaw correspondence with the DSLR's L15\textdegree/L30\textdegree/R15\textdegree/R30\textdegree{} angles, and they were therefore excluded from cross-device comparison to avoid confounding hardware effects with viewing geometry. One DSLR sample was dropped because its bounding-box annotation was invalid and prevented reliable color extraction, leaving \analyzedSampleRecords records for analysis. The public release contains \nPublicReleaseImages facial images distributed as \nDslrImages DSLR (\numDslrAngles angles), \nTabletImages tablet (\numConsumerAngles angles), and \nSmartphoneImages smartphone (\numConsumerAngles angles), implying \totalCaptureSessions capture sessions. Of these, \completeSessions sessions had a complete tri-device set covering all five analyzed regions with valid bounding-box annotations and were retained, while the remaining \droppedSessions sessions were dropped at the subject level for missing device captures, missing region annotations, or invalid bounding boxes. The DSLR provided \nDslrSamplesAllAngles samples across \numDslrAngles angles, while the tablet and smartphone each provided \nFrontPerDevice front-facing samples. All cross-device analyses use front-facing images only (\nCrossDeviceTotal samples total; \nFrontPerDevice per device). The DSLR's additional 6 non-frontal angles are used solely for per-subject angular variability by region (Supplementary Fig.~\ref{fig:region_angular}).

The matched-image structure across devices enables cross-device ICC analysis, and age-stratified, sex-balanced sampling provides demographic diversity within the cohort. The DSLR served as the reference standard. It is not equivalent to a contact spectrophotometer but provides a camera-based reference appropriate for evaluating camera-to-camera agreement in the imaging pipeline under study.

\subsection{Color Extraction}

Images were stored as 8-bit sRGB JPEG files. For each image, a region of interest (ROI) was defined by the dataset's bounding box annotations, which specify pixel coordinates $(x, y, w, h)$ for each facial region. Color values were extracted by computing the mean RGB across all pixels within the bounding box, without masking or specular highlight removal. Images were read in BGR format via OpenCV and converted to RGB prior to aggregation. The mean RGB triplet for each ROI served as the input to all subsequent color space transformations and clinical index calculations. A sensitivity analysis compared this simple full-ROI mean against a recognized robust skin-color pipeline combining YCbCr skin-pixel segmentation,\cite{ChaiNgan1999} dichromatic specular/diffuse separation,\cite{Shafer1985,Bahmani2021} and a per-channel median. The two extractions agree across the forehead, glabella, and cheeks but diverge at the chin, where skin segmentation removes shadowed and facial-hair pixels. Supplementary Fig.~\ref{fig:robust_roi} characterizes the resulting effect on cross-device reliability.

\subsection{Color Space Transformation}

Mean RGB values were treated as encoded in the canonical sRGB color space (IEC 61966-2-1\cite{IEC61966}) and decoded using the standard library implementation in \texttt{scikit-image} (\texttt{skimage.color.rgb2lab}), which applies the IEC sRGB inverse companding (a piecewise linear-and-gamma transform with knee at \srgbKnee and exponent \srgbGamma) followed by the standard linear sRGB-to-XYZ matrix transformation (which produces D65-referenced CIE XYZ directly, so no chromatic adaptation is applied) and projection to $L^*a^*b^*$ using the CIE 1931 \cieObserverDeg\textdegree{} standard observer. No per-device gamma estimate is applied, which avoids embedding device-specific assumptions in the conversion step. The inter-device comparisons therefore reflect calibration applied to canonical-sRGB-decoded inputs. The transformation process is therefore
\begin{equation*}
    \text{sRGB}_{\text{IEC 61966-2-1}} \xrightarrow{\text{IEC inverse companding}} \text{Linear RGB} \xrightarrow{\text{D65}} \text{XYZ} \xrightarrow{\text{CIE 1931 \cieObserverDeg\textdegree{}}} L^*a^*b^*\text{.}
\end{equation*}

\subsection{Color Correction Matrix}\label{sec:ccm}

Linear CCM normalization was performed using least squares, minimizing the $L_2$ norm of color error between source and DSLR reference:
\begin{equation*}
    \min_{\mathbf{A}, \mathbf{b}} \sum_{i=1}^{N} \left\| \mathbf{A} \cdot \text{RGB}_{\text{source}}^{(i)} + \mathbf{b} - \text{RGB}_{\text{ref}}^{(i)} \right\|_2^2
\end{equation*}
yielding the correction:
\begin{equation*}
    \begin{bmatrix} R_{\text{corrected}} \\ G_{\text{corrected}} \\ B_{\text{corrected}} \end{bmatrix} =
    \mathbf{A} \begin{bmatrix} R_{\text{source}} \\ G_{\text{source}} \\ B_{\text{source}} \end{bmatrix} + \mathbf{b}
\end{equation*}
where $\mathbf{A} \in \mathbb{R}^{3 \times 3}$ is the color correction matrix and $\mathbf{b} \in \mathbb{R}^3$ is the bias vector. Corrected RGB values were clipped to the valid 8-bit sRGB range [0, 255] before conversion to CIELAB to enforce a physically realisable gamut. Channel-level RMSEs reported in Fig.~\ref{fig:normalization}(b,d) are computed pre-clip to avoid deflating error magnitudes near the gamut boundary. Separate CCMs were trained for each device pair (tablet$\to$DSLR and smartphone$\to$DSLR) using matched front-facing images paired by subject and facial region. To eliminate train-test leakage in downstream reliability metrics, calibration used a \crossValidationFolds-fold cross-validation scheme with out-of-fold (OOF) prediction. Each (subject, region) pair was assigned to one of \crossValidationFolds folds by scikit-learn \texttt{GroupKFold}, which partitions on the grouping key deterministically without shuffling, so the split is reproducible without a random seed. \crossValidationFolds per-fold CCMs were fit, each using only the remaining folds, and every sample was then corrected by the CCM trained without its fold. All per-sample corrected RGB values used in $\Delta E$, ICC, and clinical-index analyses are OOF predictions. No sample is therefore corrected by a CCM that was trained on it. A supplementary analysis trained region-specific CCMs separately for each of the five facial regions using the same OOF scheme (Fig.~\ref{fig:regional_ccm}).

\subsection{Clinical Index Calculations}

The Melanin Index follows the formulation of Takiwaki,\cite{Takiwaki1998} originally defined as a logarithmic function of red-band reflectance and extended to digital imaging by Yamamoto et al.,\cite{Yamamoto2008} adapted here to CIELAB by substituting $L^*$ for reflectance:
\begin{equation*}
    \text{MI} = 100 \times \log_{10}\left(\frac{100}{L^*}\right)
\end{equation*}
The Individual Typology Angle is:\cite{Chardon1991}
\begin{equation*}
    \text{ITA} = \arctan\left(\frac{L^* - 50}{b^*}\right) \times \frac{180}{\pi}
\end{equation*}
where $L^*$ is the CIELAB lightness coordinate, $a^*$ is the red--green chromaticity coordinate, and $b^*$ is the yellow--blue chromaticity coordinate. The $a^*$ channel is reported directly as a CIELAB proxy for erythema-related redness.

\subsection{Statistics and Reproducibility}

Inter-device agreement was assessed using the Intraclass Correlation Coefficient ICC(3,1), the two-way mixed-effects, consistency-based Shrout--Fleiss Model~3,\cite{Shrout1979,Koo2016} treating devices as fixed effects and subject--region pairs as random targets. Pooling the five facial regions of each subject into the target set treats within-subject regional variation as between-target variance, yielding a conservative pooled estimate of within-region agreement. Model~3 was chosen over Model~2 because the three devices constitute a fixed, exhaustive set rather than a random sample from a larger device population. The consistency form was chosen over absolute agreement because systematic mean differences between devices are already removed by the additive bias term $\mathbf{b}$ of the per-device CCM normalization. The single-measure form ICC(3,1) was chosen over the average-measure ICC(3,k) because clinical measurements are obtained from a single device. The analysis comprised n=\nFrontPerDevice matched measurements per device (\numSubjects subjects $\times$ \numFacialRegionsAnalyzed regions across \numDevices devices). ICC(3,1) quantifies measurement reliability by comparing between-target variance to total variance, where higher values indicate that device differences are small relative to true between-target differences. 95\% confidence intervals for ICC were computed using the $F$-distribution method.\cite{Shrout1979} ICC values were interpreted as poor ($<$0.50), moderate (0.50--0.75), good (0.75--0.90), or excellent ($>$0.90) based on established thresholds.\cite{Portney2015}

Bland-Altman analysis characterized device agreement by plotting measurement differences against their means, with 95\% limits of agreement (approximately $\pm$\loaSdMultiplier SD, with $t$-based endpoints at the per-sample $n$) indicating the range within which most device discrepancies fall.\cite{Bland1986} No pre-specified clinically meaningful limits of agreement were set because no established thresholds exist for individual CIELAB channel differences in dermatological imaging. The sample size (n=\nFrontPerDevice per device pair) exceeds the recommended minimum of \blandAltmanMinSamples for stable Bland-Altman estimates.

Pearson correlation coefficients were computed to assess linear association between consumer-device and DSLR clinical index values after CCM normalization (Fig.~\ref{fig:correlation}). Pearson $r$ was chosen because the clinical indices are continuous and approximately linearly related across devices after linear CCM correction. Spearman rank correlation was not required because the linear calibration preserves monotonic relationships and the analysis focuses on linear agreement rather than ordinal association. Analysis of variance (ANOVA) assessed factor influences on $\Delta E$ by testing whether group means differ significantly. A single factorial ANOVA model with Type~III sums of squares partitioned variance jointly across five categorical predictors (Facial Region, Source Device, Sex, Skin Condition, Age Group). Each factor's $F$-test and effect size therefore reflect its marginal contribution after adjusting for the other four. Effect sizes were quantified using eta-squared ($\eta^2$), representing the proportion of total variance explained by each factor, where values of 0.01, 0.06, and 0.14 were interpreted as small, medium, and large effects, respectively.\cite{Cohen1988}

All analyses used Python 3.10 with numpy ($\geq$1.24.0), scipy ($\geq$1.10.0), opencv-python ($\geq$4.7.0), scikit-learn ($\geq$1.2.0), pingouin ($\geq$0.5.3), and statsmodels ($\geq$0.14.0). Values in parentheses are minimum package versions. Significance was set at $\alpha = \alphaSig$ with Bonferroni correction for the \numAnovaTests simultaneous ANOVA tests (adjusted threshold $\alpha = \alphaBonferroni$). All statistical tests were two-sided. Exact $F$-statistics, $p$-values, and effect sizes ($\eta^2$) for all ANOVA tests are reported in Table~\ref{tab:anova}.

\section{Results}\label{sec:results}

\subsection{Clinical Unusability of Uncalibrated Images}

\begin{figure}[ht]
\centering
\includegraphics[width=0.9\textwidth]{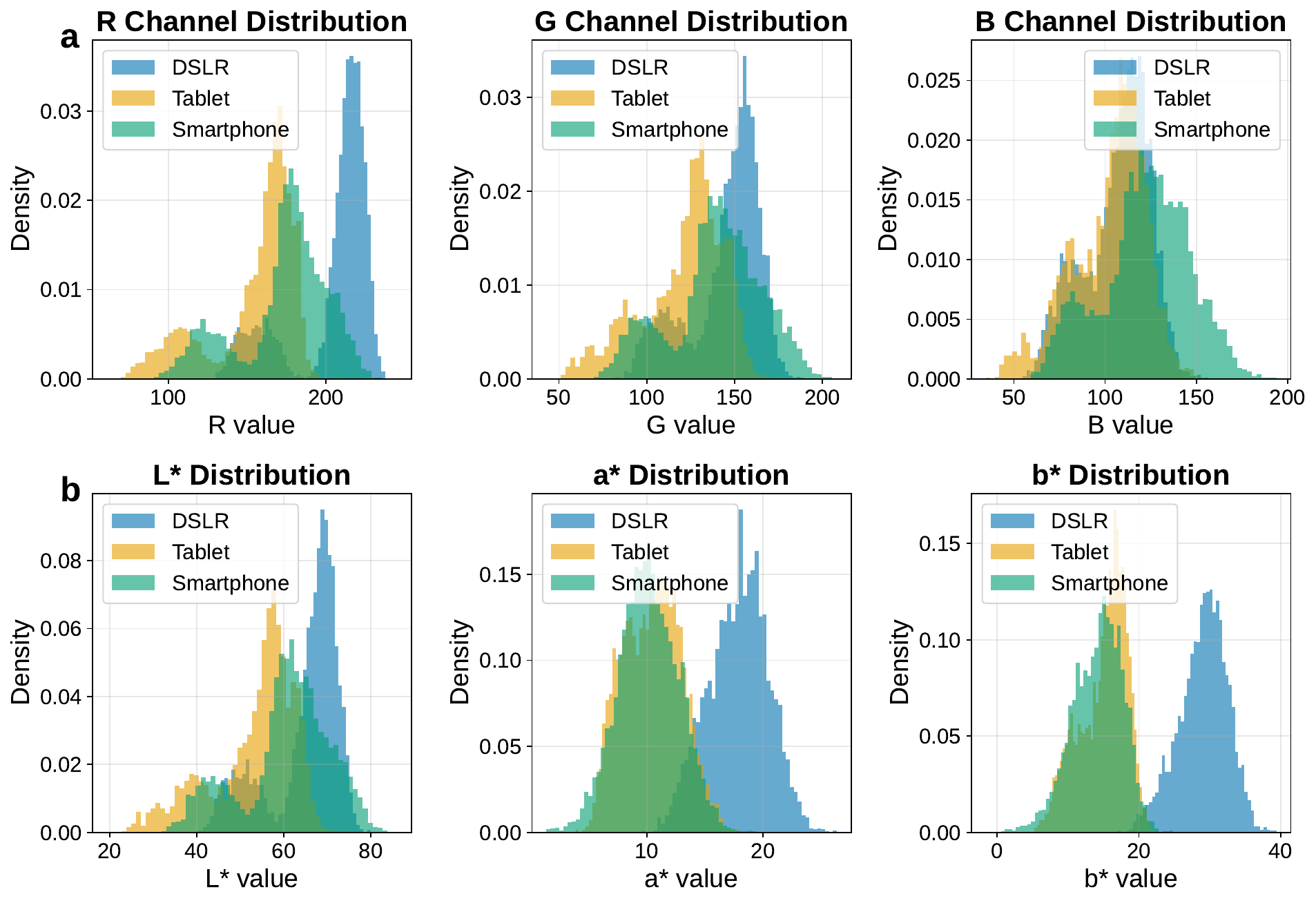}
\caption{Consumer devices show systematic color rendering compression relative to the digital single-lens reflex (DSLR) reference in both RGB and Commission Internationale de l'\'Eclairage L*a*b* (CIELAB) spaces (n=\nFrontPerDevice per device, front-facing only). a) RGB channel distributions showing device-specific color rendering characteristics. Histograms show normalized density for DSLR (blue), tablet (orange), and smartphone (green). b) CIELAB $L^*a^*b^*$ distributions demonstrating lightness and chromaticity variations across devices, with blue, orange, and green for DSLR, tablet, and smartphone, respectively.}\label{fig:color_response}
\end{figure}

Consumer devices exhibited systematic underexposure relative to the digital single-lens reflex (DSLR) reference (mean $L^* = \meanLstarTablet$ for tablet and $\meanLstarSmartphone$ for smartphone, versus $\meanLstarDslr$ for DSLR; n=\nFrontPerDevice front-facing images per device, \numSubjects subjects $\times$ \numFacialRegionsAnalyzed regions) and compressed chromaticity values, as visualized in Fig.~\ref{fig:color_response}. Mean $a^*$ values were attenuated ($\meanAstarSmartphone$--$\meanAstarTablet$ versus $\meanAstarDslr$) and mean $b^*$ values were attenuated ($\meanBstarSmartphone$--$\meanBstarTablet$ versus $\meanBstarDslr$).
Table~\ref{tab:color_shift} presents inter-device color differences. Mean $\Delta E$ values of $\deltaEMeanTabletDslr$ (tablet) and $\deltaEMeanSmartphoneDslr$ (smartphone) versus DSLR exceeded the clinical threshold ($\Delta E < \deltaEThreshold$) by \deltaEFoldExceedSmartphone- to \deltaEFoldExceedTablet-fold. No consumer-device-versus-DSLR image pair achieved $\Delta E < \deltaEThreshold$. The minimum observed $\Delta E$ was \deltaEMinSmartphoneDslr (smartphone vs.\ DSLR), a margin robust to alternative region-of-interest extraction methods (Supplementary Fig.~\ref{fig:robust_roi}).

\begin{figure}[ht]
\centering
\includegraphics[width=0.95\textwidth]{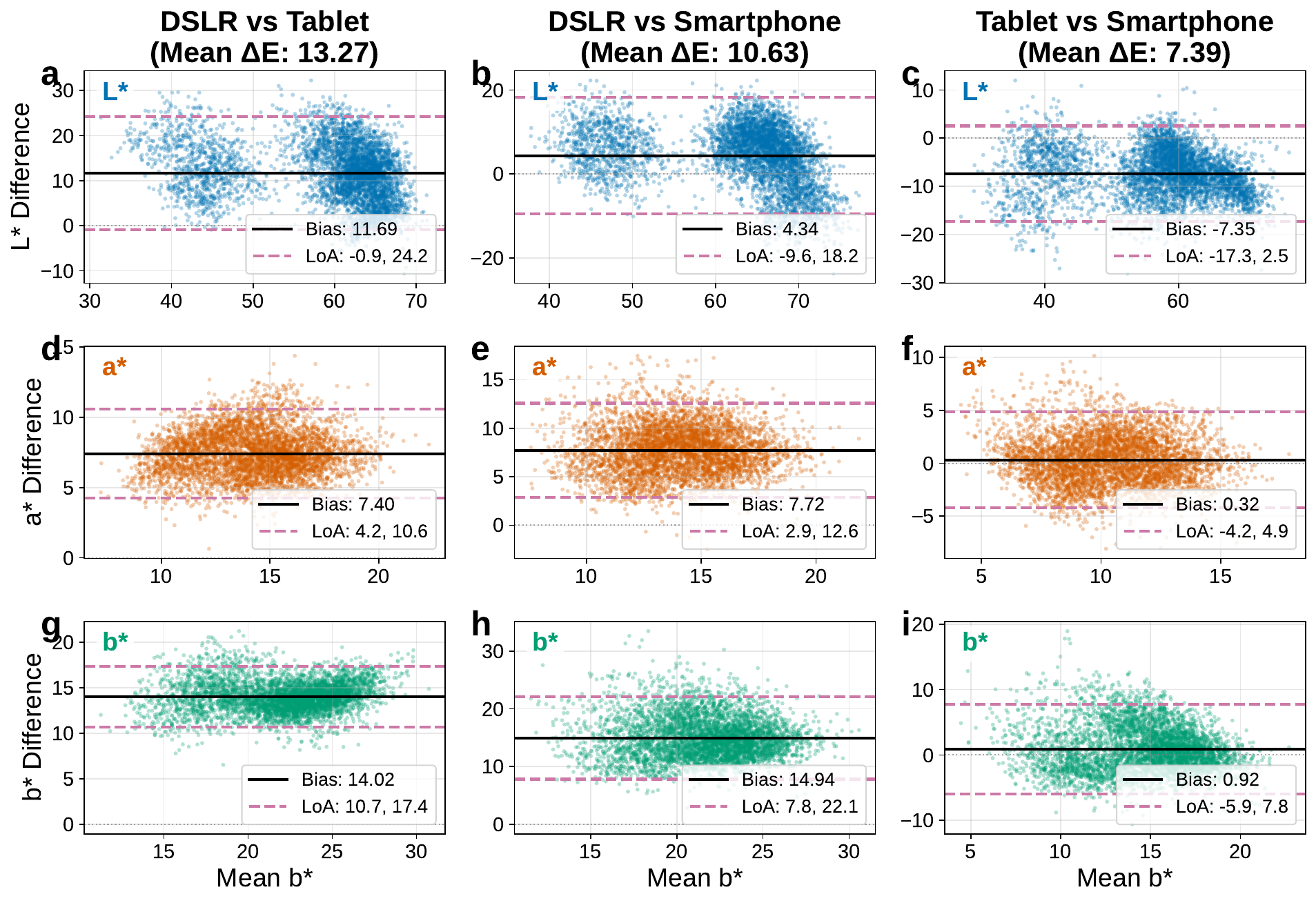}
\caption{Bland-Altman analysis of inter-device disagreement on each Commission Internationale de l'\'Eclairage L*a*b* (CIELAB) channel, showing affine offsets across the cohort range that a linear Color Correction Matrix can correct (n=\nFrontPerDevice matched pairs per device pair). Each panel plots individual measurements as colored scatter points, with blue for $L^*$, vermilion for $a^*$ (red--green axis), and bluish green for $b^*$ (yellow--blue axis). Black solid lines indicate mean bias. Pink dashed lines show 95\% limits of agreement (approximately $\pm 1.96$ SD; $t$-based endpoints). Grey dotted lines mark zero difference. DSLR, digital single-lens reflex; SD, standard deviation.}\label{fig:bland_altman}
\end{figure}

Bland-Altman analysis (Fig.~\ref{fig:bland_altman}) shows that the $L^*$ disagreement, although the largest in absolute magnitude, follows a consistent linear trend across the range of facial $L^*$ values. The $a^*$ and $b^*$ chromaticity channels show analogous affine offsets at smaller magnitudes. The structure is fully addressable by the affine form of a linear color correction matrix, motivating the standard CCM treatment that follows.

\begin{table}[ht]
\caption{Inter-device CIEDE2000 color differences ($\Delta E$) before Color Correction Matrix normalization (n=\nFrontPerDevice matched front-facing image pairs per device combination across \numSubjects subjects and 5 facial regions). Clinical acceptability threshold is $\Delta E < \deltaEThreshold$.}\label{tab:color_shift}
\centering
\begin{tabular}{lcccc}
\toprule
\textbf{Device Pair} & \textbf{Mean} & \textbf{Std} & \textbf{Median} & \textbf{95th \%ile} \\
\midrule
DSLR vs Tablet & \deltaEMeanTabletDslr & \deltaEStdTabletDslr & \deltaEMedianTabletDslr & \deltaEPctileTabletDslr \\
DSLR vs Smartphone & \deltaEMeanSmartphoneDslr & \deltaEStdSmartphoneDslr & \deltaEMedianSmartphoneDslr & \deltaEPctileSmartphoneDslr \\
Tablet vs Smartphone & \deltaEMeanTabletSmartphone & \deltaEStdTabletSmartphone & \deltaEMedianTabletSmartphone & \deltaEPctileTabletSmartphone \\
\bottomrule
\end{tabular}
\end{table}

\subsection{CCM Normalization: Colorimetric Success}\label{sec:ccm_success}

\begin{figure}[ht]
\centering
\includegraphics[width=0.85\textwidth]{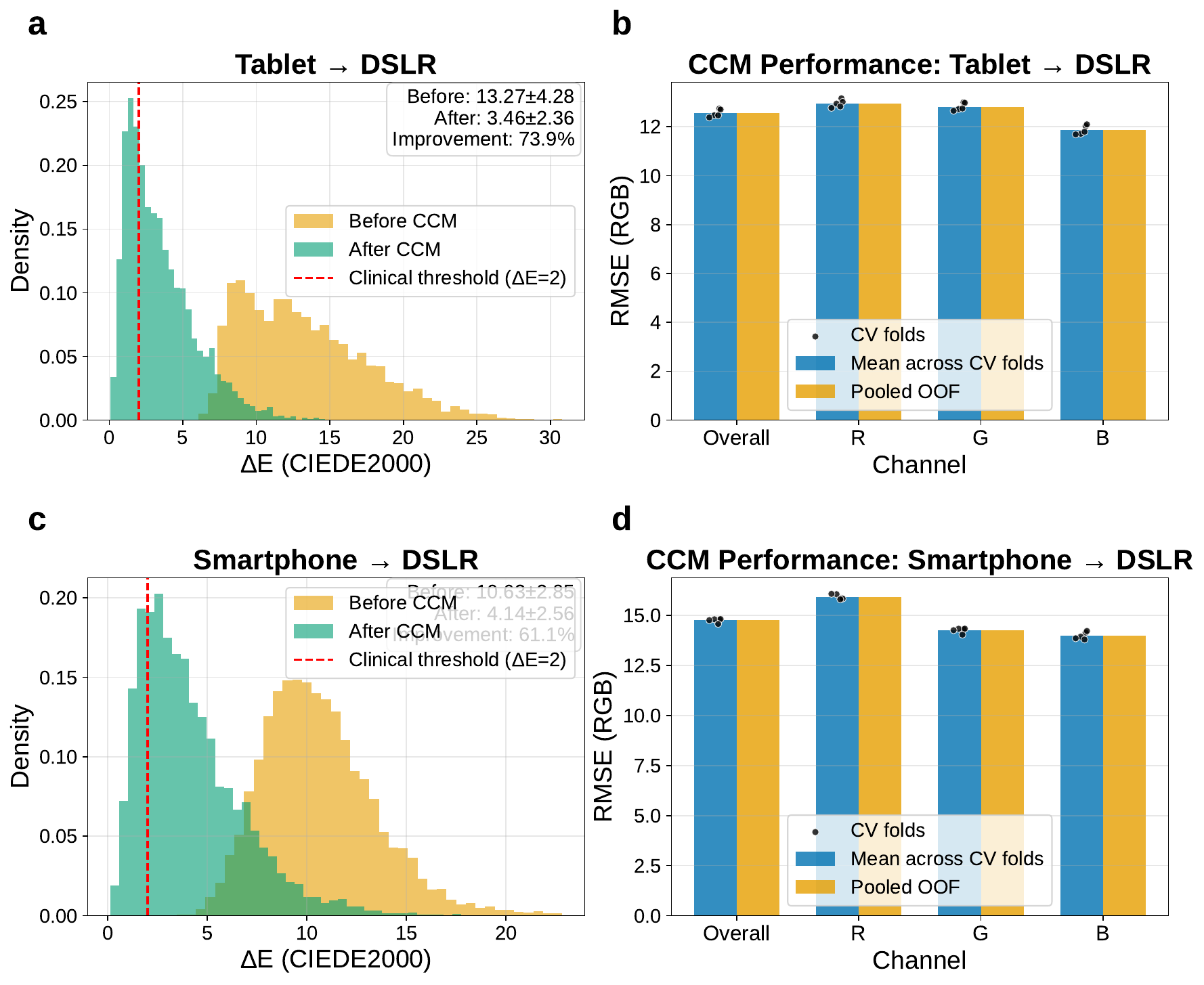}
\caption{CCM normalization reduces inter-device color differences substantially (n=\nFrontPerDevice out-of-fold samples per transformation). a,~c) $\Delta E$ (CIEDE2000) distributions before (orange histograms) and after (green histograms) out-of-fold (OOF) calibration for tablet$\to$digital single-lens reflex (DSLR) (a) and smartphone$\to$DSLR (c); red dashed lines indicate the clinical acceptability threshold ($\Delta E = \deltaEThreshold$); inset text boxes report mean $\pm$ SD and percent improvement. b,~d) Root mean squared error (RMSE) per RGB channel: mean across the five CV fold validations (blue bars) versus the pooled OOF prediction across all samples (orange bars), confirming stable performance. SD, standard deviation.}\label{fig:normalization}
\end{figure}

Linear CCM reduced inter-device color differences by \ccmReductionLow--\ccmReductionHigh\% (Fig.~\ref{fig:normalization}). All calibration performance values below are out-of-fold (OOF) predictions from \crossValidationFolds-fold cross-validation, in which every sample is corrected by the CCM trained on the remaining folds. The OOF scheme affects only the post-calibration $\Delta E$; the pre-calibration baselines used for the improvement percentages equal the raw means in Table~\ref{tab:color_shift} exactly. Tablet showed \ccmTabletImprovementPct\% improvement (mean $\pm$ SD $\Delta E$ from $\deltaEMeanTabletDslr \pm \deltaEStdTabletDslr$ to $\deltaETabletPostCcmMean \pm \deltaETabletPostCcmStd$). Smartphone showed \ccmSmartphoneImprovementPct\% improvement (from $\deltaEMeanSmartphoneDslr \pm \deltaEStdSmartphoneDslr$ to $\deltaESmartphonePostCcmMean \pm \deltaESmartphonePostCcmStd$). Post-normalization median $\Delta E$ approached the clinical acceptability threshold for tablet (\deltaETabletPostCcmMedian) but remained above it for both devices, indicating that linear correction reduces but does not eliminate the inter-device color gap. The clinical-relevance question therefore depends on the inter-device reliability of the derived clinical indices, which is examined next.

\subsection{Inter-Device Reliability of Clinical Indices Under Standard CCM}\label{sec:reliability}

\begin{figure}[ht]
\centering
\includegraphics[width=0.85\textwidth]{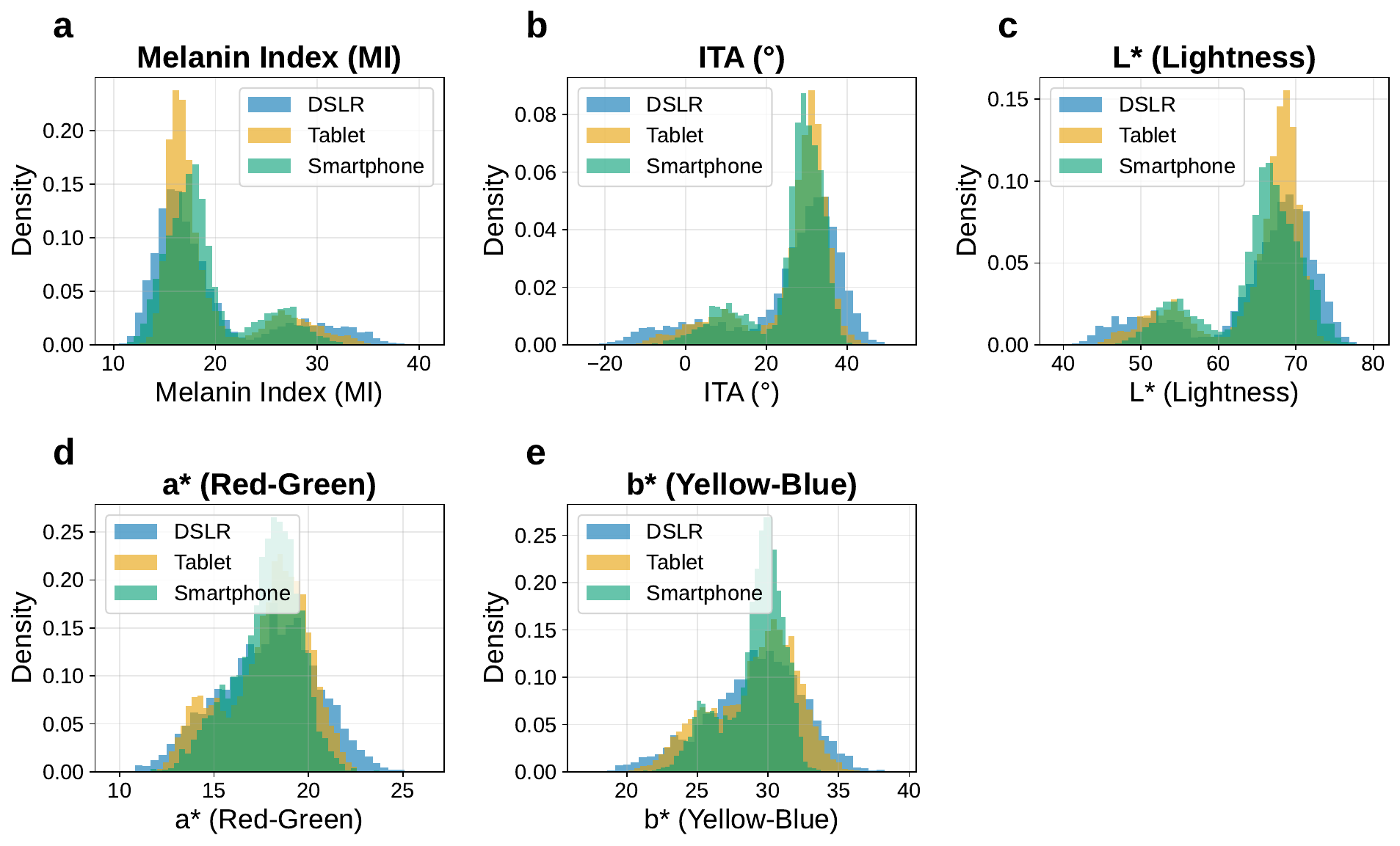}
\caption{Clinical index distributions after Color Correction Matrix normalization (n=\nFrontPerDevice per device). Melanin Index and Individual Typology Angle distributions overlap across devices, consistent with the good inter-device agreement reported in Table~\ref{tab:clinical_icc}. Normalized density histograms show distributions for digital single-lens reflex (DSLR, blue), tablet (orange), and smartphone (green). a) Melanin Index (MI). b) Individual Typology Angle (ITA). c) $L^*$ (lightness). d) $a^*$ (red--green). e) $b^*$ (yellow--blue). CCM, Color Correction Matrix; CIELAB, Commission Internationale de l'\'Eclairage L*a*b*.}\label{fig:clinical_indices}
\end{figure}

\begin{figure}[ht]
\centering
\includegraphics[width=0.9\textwidth]{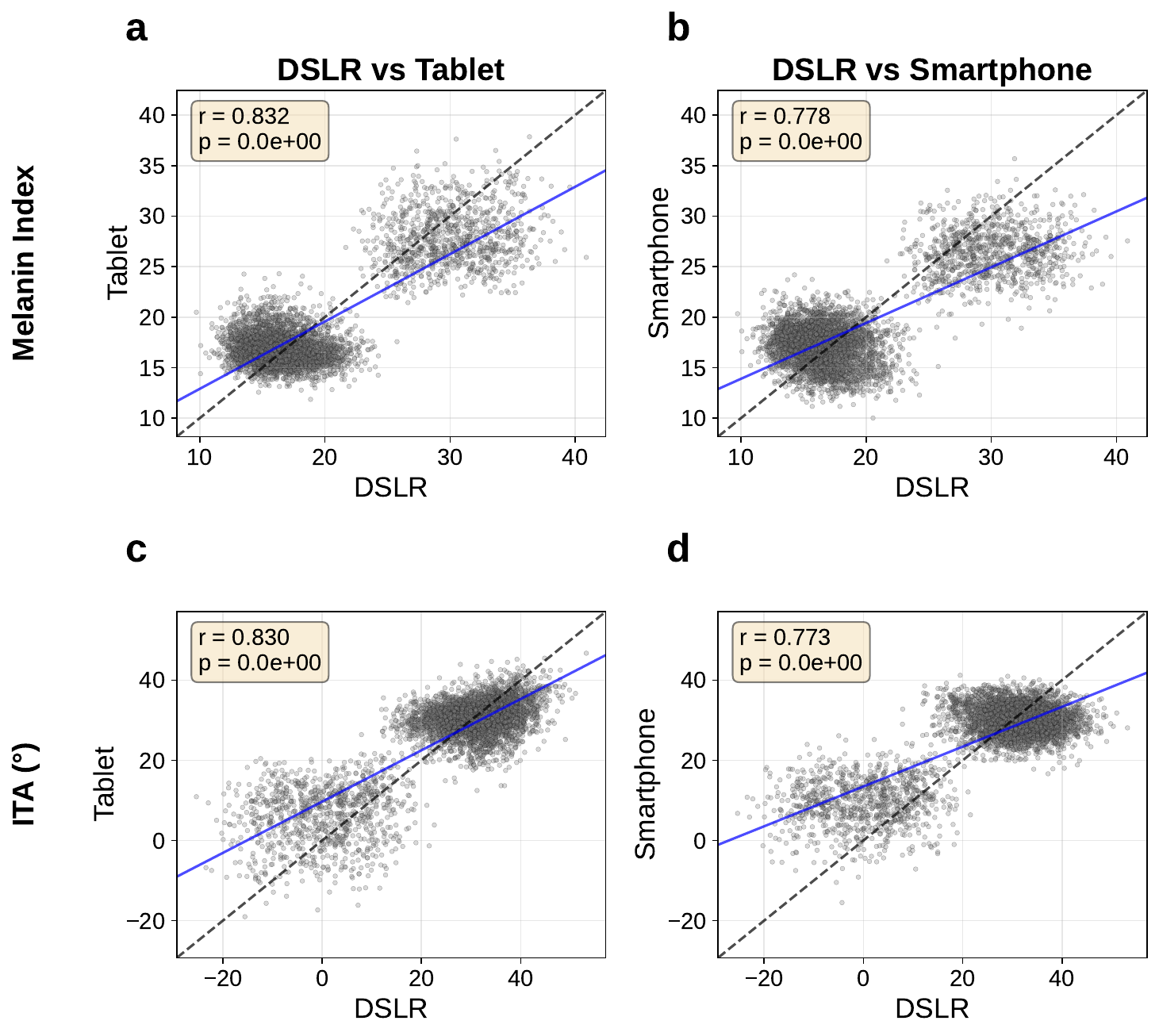}
\caption{Per-sample agreement between consumer devices and the digital single-lens reflex (DSLR) reference after Color Correction Matrix (CCM) normalization. Scatter plots (grey circles) show Melanin Index (MI) and Individual Typology Angle (ITA) for tablet versus DSLR (left column) and smartphone versus DSLR (right column), with dashed diagonal identity lines and solid blue regression lines (n=\nFrontPerDevice per device pair). Pearson correlation coefficients ($r$) and $p$-values (two-tailed) are annotated in each panel. Both indices show tight scatter along the identity line, consistent with their good inter-device ICC (Table~\ref{tab:clinical_icc}). CIELAB, Commission Internationale de l'\'Eclairage L*a*b*.}\label{fig:correlation}
\end{figure}

Both clinical indices reached the good reliability band using standard global CCM (Table~\ref{tab:clinical_icc}, Fig.~\ref{fig:clinical_indices}, and Fig.~\ref{fig:correlation}). ICC(3,1) was computed across all three devices (n=\nFrontPerDevice per device, pooling the five facial regions per subject). The pooled metric reflects current clinical practice, in which a single CCM is applied uniformly across facial regions.\cite{Hong2001} Melanin Index reached ICC = \iccMelaninGlobal and ITA reached ICC = \iccItaGlobal, both above the 0.75 good-reliability threshold.\cite{Portney2015}

\begin{table}[ht]
\caption{Intraclass Correlation Coefficients (ICC(3,1), two-way mixed-effects model) for clinical indices and Commission Internationale de l'\'Eclairage L*a*b* (CIELAB) channels after Color Correction Matrix (CCM) normalization (n=\nFrontPerDevice matched samples per device, 3 devices, pooled across five facial regions). ICC interpretation thresholds following Portney and Watkins,\cite{Portney2015} where poor is below 0.50, moderate is 0.50 to 0.75, good is 0.75 to 0.90, and excellent is above 0.90.}\label{tab:clinical_icc}
\centering
\begin{tabular}{lccc}
\toprule
\textbf{Index/Channel} & \textbf{ICC(3,1)} & \textbf{95\% CI} & \textbf{Interpretation} \\
\midrule
Melanin Index & \iccMelaninGlobal & \iccMelaninGlobalCI & Good \\
ITA & \iccItaGlobal & \iccItaGlobalCI & Good \\
$L^*$ & \iccLstarGlobal & \iccLstarGlobalCI & Good \\
$a^*$ & \iccAstarGlobal & \iccAstarGlobalCI & Moderate \\
$b^*$ & \iccBstarGlobal & \iccBstarGlobalCI & Moderate \\
\bottomrule
\end{tabular}
\end{table}

Index reliability tracks the underlying CIELAB channels. The Melanin Index inherits the good agreement of $L^*$ (ICC = \iccLstarGlobal). ITA depends on both $L^*$ and the moderate-agreement $b^*$ channel (ICC = \iccBstarGlobal). The resulting ITA agreement (\iccItaGlobal) thus lies between the two and clears the good-reliability threshold.

This good agreement is not an artifact of the simple full-ROI color extraction, although the ICC value itself depends on it. A robust skin-pixel segmentation pipeline lowers the pre-calibration pooled cross-device ICC from \iccRobustRoiMiSimple to \iccRobustRoiMiRobust (Melanin Index) and from \iccRobustRoiItaSimple to \iccRobustRoiItaRobust (ITA) (Supplementary Fig.~\ref{fig:robust_roi}). This decline is driven by a \varBetweenItaReductionPct--\varBetweenMiReductionPct\% reduction in pooled between-subject variance across the (subject, region) targets, not by worse inter-device agreement. The chin is the region responsible, because its segmentation discards shadowed and facial-hair pixels that the other four regions lack. The cross-device variance component is unchanged for the Melanin Index and smaller for ITA under the robust extraction. The consumer devices therefore agree at least as well under the robust pipeline, even though its ICC is lower. The good-reliability band reflects pooled cohort heterogeneity rather than inter-device measurement error.

Distributional signatures match the ICC pattern (Fig.~\ref{fig:clinical_indices}). Melanin Index distributions (Fig.~\ref{fig:clinical_indices}(a)) and ITA distributions (Fig.~\ref{fig:clinical_indices}(b)) overlap closely across the three devices, matching the good ICC values in Table~\ref{tab:clinical_icc}. Both indices retain a bimodal shape inside every device, with a secondary mode offset from the dominant mode along the index axis, and the same bimodal shape appears on all three devices rather than only on the DSLR. Per-sample scatter plots (Fig.~\ref{fig:correlation}) show the same two-cluster pattern along the identity line for both indices. Consumer-device values therefore match the DSLR reference sample by sample, not only on average. A within-cohort subpopulation effect therefore survives global CCM, identifying a residual variance source that uniform alignment cannot remove and motivating the variance decomposition that follows.

\subsection{Sources of Variance in Cross-Device Color Differences}\label{sec:variance}

\begin{figure}[ht]
\centering
\includegraphics[width=0.9\textwidth]{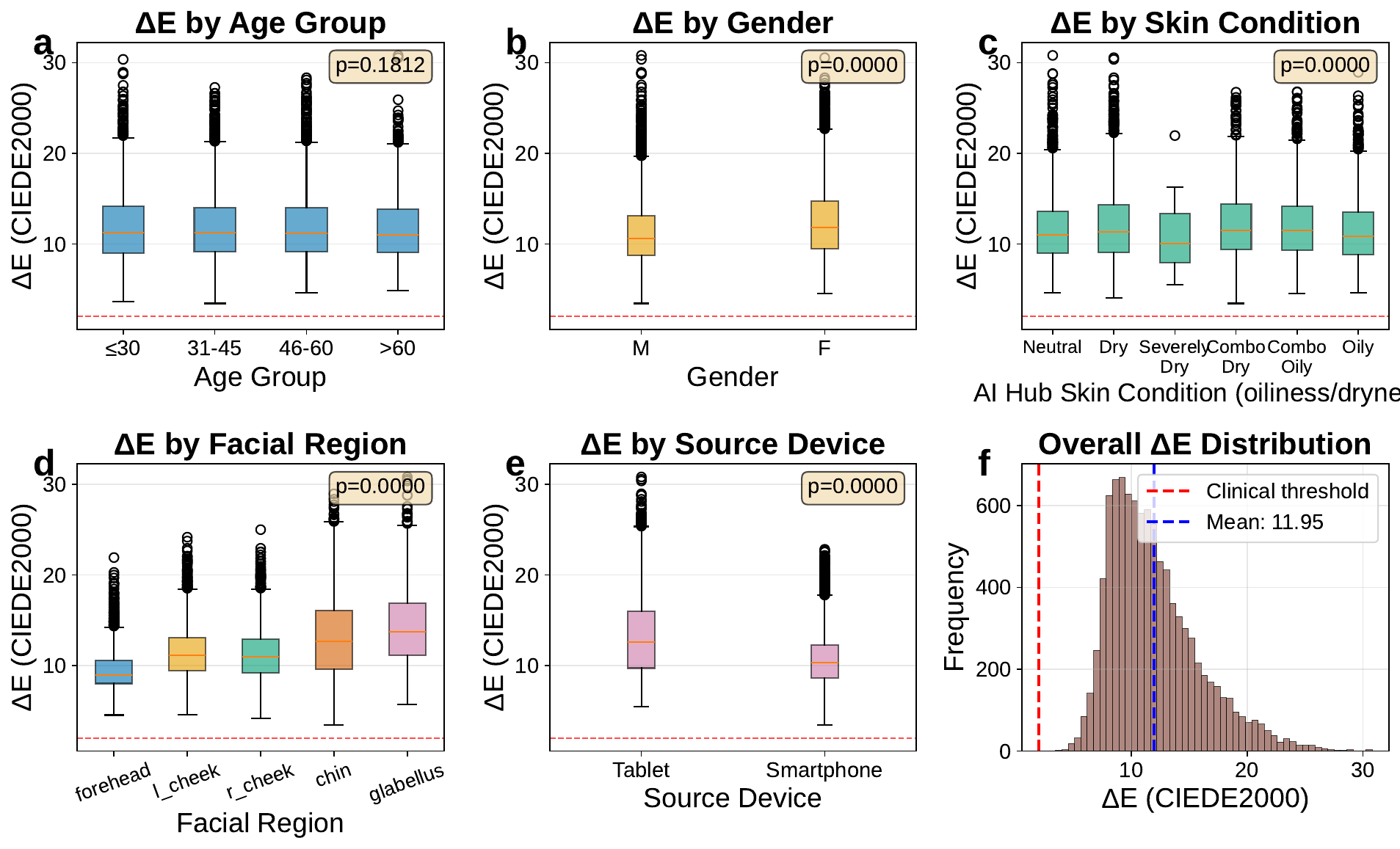}
\caption{Stratified analysis of pre-normalization DSLR-vs-consumer color differences ($\Delta E$, CIEDE2000) across demographic, anatomical, and device factors (n=\nAnovaPairs total matched pairs). Box plots show median (horizontal line), interquartile range (IQR; box), whiskers extending to 1.5$\times$ IQR, and individual outliers (circles). Red dashed lines indicate the clinical acceptability threshold ($\Delta E = \deltaEThreshold$). a) Age groups (blue boxes). b) Sex (orange boxes). c) AI Hub skin condition class (sebum/moisture phenotype: neutral, dry, severely dry, combination dry, combination oily, oily; green boxes). d) Facial region (blue, orange, green, vermilion, and pink boxes for forehead, left cheek, right cheek, chin, and glabella, respectively). e) Source device (pink boxes). f) Overall $\Delta E$ distribution (brown histogram) with blue dashed line at the mean. Factorial ANOVA (Type~III sums of squares) $p$-values are shown in each panel; see Table~\ref{tab:anova} for $F$-statistics and effect sizes. DSLR, digital single-lens reflex.}\label{fig:robustness}
\end{figure}

A factorial ANOVA on pre-normalization $\Delta E$ identified anatomical region as the dominant variance source (\anovaEtaRegionPct\% of variance, $\eta^2 = \anovaEtaRegion$, $p < 10^{-300}$), approximately \anovaRegionDeviceRatio times larger than source device ($\eta^2 = \anovaEtaDevice$; Table~\ref{tab:anova} and Fig.~\ref{fig:robustness}). Demographic factors contributed far less variance. Sex showed a small effect ($\eta^2 = \anovaEtaSex$), and the AI Hub skin-condition class ($\eta^2 = \anovaEtaSkin$, $p = \anovaPSkin$; not significant at the Bonferroni-corrected threshold) and age group ($\eta^2 = \anovaEtaAge$) were negligible. Region and source device are the two factors that any practical calibration must address.

Tablet-versus-smartphone color differences (mean $\Delta E = \deltaEMeanTabletSmartphone$; Table~\ref{tab:color_shift}) were \consumerVsCrossConsumerRatioLow--\consumerVsCrossConsumerRatioHigh times smaller than either device's deviation from the DSLR reference, indicating that the two consumer pipelines share a broadly similar distortion profile relative to the professional reference. The regional structure of that distortion is itself device-dependent, however (Supplementary Fig.~\ref{fig:dE_by_region}): tablet-vs-DSLR $\Delta E$ varies markedly across regions, whereas smartphone-vs-DSLR $\Delta E$ is comparatively uniform.

The regional contribution reflects anatomical structure rather than measurement noise. Per-subject color variability across DSLR viewing angles varies by region, with chin highest and forehead lowest (Supplementary Fig.~\ref{fig:region_angular}). The DSLR reference itself is stable across viewing angles, with multi-angle self-agreement ICC of \iccDslrCeilingLow--\iccDslrCeilingHigh across indices (Supplementary Fig.~\ref{fig:dslr_ceiling}) substantially exceeding cross-device ICC values and confirming the DSLR as an adequate reference. The magnitude of the regional contribution motivates the region-specific calibration evaluated in Section~\ref{sec:regional}.

\begin{table}[ht]
\caption{Factorial analysis of variance (ANOVA) with Type~III sums of squares for factors influencing pre-normalization inter-device color differences ($\Delta E$, CIEDE2000; n=\nAnovaPairs matched device pairs across 2 consumer-device-vs-DSLR pairs by \nFrontPerDevice samples). All five factors enter the OLS model jointly. Each factor's $F$-test and effect size therefore quantify its marginal contribution after adjusting for the other four. Effect size labels follow Cohen (1988) thresholds (negligible $\eta^2 < 0.01$, small $\eta^2 \geq 0.01$, medium $\eta^2 \geq 0.06$, large $\eta^2 \geq 0.14$). Bonferroni-corrected significance threshold $\alpha = \alphaBonferroni$ for the five simultaneous $F$-tests.}\label{tab:anova}
\centering
\begin{tabular}{lcccc}
\toprule
\textbf{Factor} & \textbf{F-statistic} & \textbf{p-value} & \textbf{$\eta^2$} & \textbf{Effect Size} \\
\midrule
Facial Region & \anovaFRegion & $\anovaPRegion$ & \anovaEtaRegion & Large \\
Source Device & \anovaFDevice & $\anovaPDevice$ & \anovaEtaDevice & Medium \\
Sex & \anovaFSex & $\anovaPSex$ & \anovaEtaSex & Small \\
Skin Condition (oiliness/dryness) & \anovaFSkin & $\anovaPSkin$ & \anovaEtaSkin & Negligible \\
Age Group & \anovaFAge & $\anovaPAge$ & \anovaEtaAge & Negligible \\
\bottomrule
\end{tabular}
\end{table}

\subsection{Region-Specific Calibration Achieves Excellent Inter-Device Reliability}\label{sec:regional}

\begin{figure}[ht]
\centering
\includegraphics[width=0.9\textwidth]{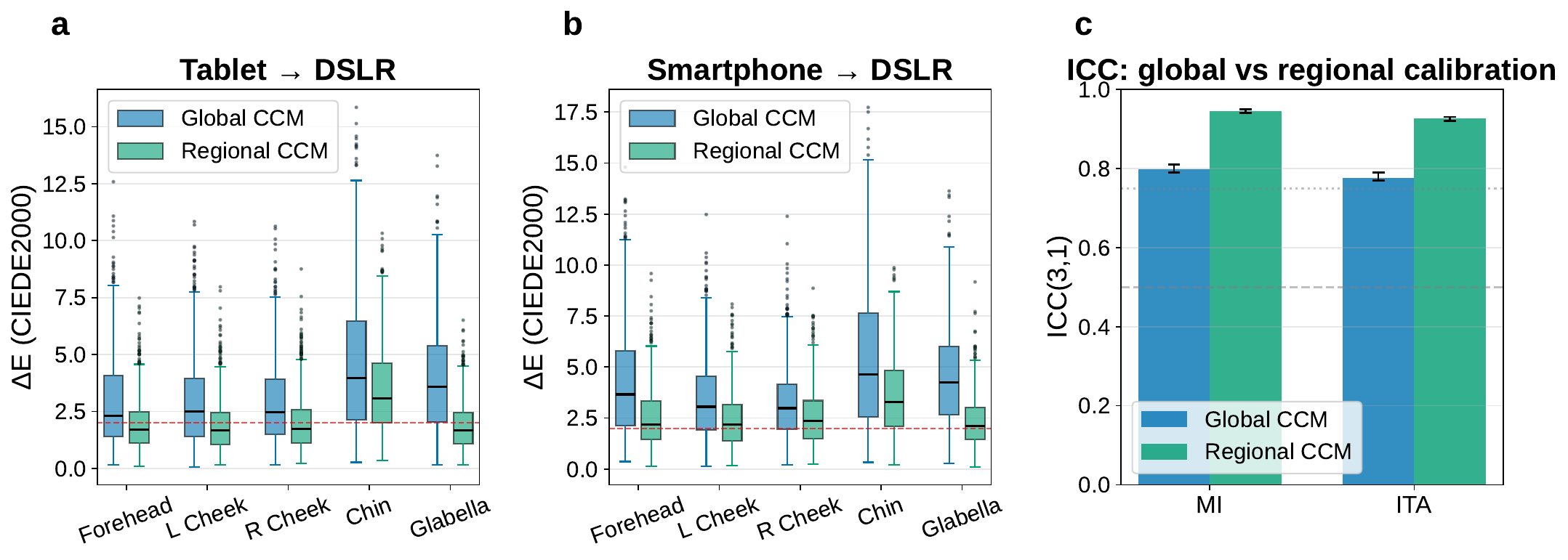}
\caption{Region-specific Color Correction Matrices (CCMs) reduce post-calibration color differences and raise clinical-index ICC into the excellent band. a,~b) Mean $\Delta E$ (CIEDE2000) per facial region, computed as out-of-fold (OOF) predictions from \crossValidationFolds-fold cross-validation, for tablet to digital single-lens reflex (DSLR) (a) and smartphone to DSLR (b) transformations, comparing global CCM (blue bars) with region-specific CCMs (green bars). Red dashed lines indicate the clinical acceptability threshold ($\Delta E = \deltaEThreshold$). c) ICC(3,1) comparison for clinical indices using global versus regional calibration. Region-specific CCMs raise ITA ICC from \iccItaGlobal (good) to \iccItaRegional (excellent) and Melanin Index ICC from \iccMelaninGlobal (good) to \iccMelaninRegional (excellent). Per-region $\Delta E$ values before and after regional calibration are reported in Table~\ref{tab:regional_ccm_summary}. ICC, intraclass correlation coefficient.}\label{fig:regional_ccm}
\end{figure}

Region-specific CCMs trained separately for each of the five facial regions raised both clinical indices into the excellent reliability band (Fig.~\ref{fig:regional_ccm}). Per-region $\Delta E$ decreased by \regionalCcmReductionPctTablet\% (tablet) and \regionalCcmReductionPctSmartphone\% (smartphone) relative to the global CCM baseline (mean $\Delta E$ from \deltaETabletGlobalCcmRegionalMean to \deltaETabletRegionalCcmRegionalMean and from \deltaESmartphoneGlobalCcmRegionalMean to \deltaESmartphoneRegionalCcmRegionalMean, respectively, Fig.~\ref{fig:regional_ccm}(a,b)). ITA ICC rose from \iccItaGlobal (good) using global CCM to \iccItaRegional (excellent) using regional CCMs, and Melanin Index ICC rose from \iccMelaninGlobal (good) to \iccMelaninRegional (excellent) (Fig.~\ref{fig:regional_ccm}(c)). The chin retains the largest absolute residual $\Delta E$ post-regional CCM ($\deltaEChinTabletRegional$ for tablet versus $\deltaEOtherRegionsTabletLow$--$\deltaEOtherRegionsTabletHigh$ for the other four regions, and $\deltaEChinSmartphoneRegional$ for smartphone versus $\deltaEOtherRegionsSmartphoneLow$--$\deltaEOtherRegionsSmartphoneHigh$), reflecting a region-specific residual that uniform-within-region calibration does not remove. Smartphone post-regional $\Delta E$ values remain above the $\Delta E < \deltaEThreshold$ clinical threshold in all five regions, despite a \deltaERcheekSmartphoneRegionalReductionPct--\deltaEGlabellaSmartphoneRegionalReductionPct\% reduction relative to global CCM (Table~\ref{tab:regional_ccm_summary}). Regional CCM nevertheless raises rank-order agreement into the excellent ICC band, because ICC measures subject rank preservation under the calibrated transform rather than absolute color difference at each sample.

The region-specific result identifies anatomical region as the dominant remaining target for improvement after standard calibration. Combined with the ANOVA finding that region edges the source device as the largest pre-calibration variance contributor ($\eta^2 = \anovaEtaRegionAbstract$ versus $\anovaEtaDeviceAbstract$, Section~\ref{sec:variance}), the practical implication for smartphone deployments is direct. A region-aware CCM that selects its correction matrix based on facial-landmark detection of anatomical regions raises ITA into the excellent reliability band on the same hardware that achieves only good reliability using uniform calibration. The same regional calibration reduces ITA-based skin-type misclassification, the clinical-deployment translation of the ICC gain evaluated next in Section~\ref{sec:misclassification}.

\begin{table}[ht]
\caption{Per-region mean $\Delta E$ (CIEDE2000) using global versus region-specific Color Correction Matrix (CCM) normalization, out-of-fold predictions from \crossValidationFolds-fold cross-validation (n=\nFrontPerDevice matched samples per device pair). Reduction is $(\Delta E_\mathrm{global}-\Delta E_\mathrm{regional})/\Delta E_\mathrm{global}\times 100$.}\label{tab:regional_ccm_summary}
\centering
\begin{tabular}{lcccccc}
\toprule
 & \multicolumn{3}{c}{Tablet vs DSLR} & \multicolumn{3}{c}{Smartphone vs DSLR} \\
\cmidrule(lr){2-4}\cmidrule(lr){5-7}
Region & Global & Regional & Reduction (\%) & Global & Regional & Reduction (\%) \\
\midrule
Forehead    & \deltaEForeheadTabletGlobal & \deltaEForeheadTabletRegional & \deltaEForeheadTabletRegionalReductionPct & \deltaEForeheadSmartphoneGlobal & \deltaEForeheadSmartphoneRegional & \deltaEForeheadSmartphoneRegionalReductionPct \\
Left cheek  & \deltaELcheekTabletGlobal & \deltaELcheekTabletRegional & \deltaELcheekTabletRegionalReductionPct & \deltaELcheekSmartphoneGlobal & \deltaELcheekSmartphoneRegional & \deltaELcheekSmartphoneRegionalReductionPct \\
Right cheek & \deltaERcheekTabletGlobal & \deltaERcheekTabletRegional & \deltaERcheekTabletRegionalReductionPct & \deltaERcheekSmartphoneGlobal & \deltaERcheekSmartphoneRegional & \deltaERcheekSmartphoneRegionalReductionPct \\
Chin        & \deltaEChinTabletGlobal & \deltaEChinTabletRegional & \deltaEChinTabletRegionalReductionPct & \deltaEChinSmartphoneGlobal & \deltaEChinSmartphoneRegional & \deltaEChinSmartphoneRegionalReductionPct \\
Glabella    & \deltaEGlabellaTabletGlobal & \deltaEGlabellaTabletRegional & \deltaEGlabellaTabletRegionalReductionPct & \deltaEGlabellaSmartphoneGlobal & \deltaEGlabellaSmartphoneRegional & \deltaEGlabellaSmartphoneRegionalReductionPct \\
\bottomrule
\end{tabular}
\end{table}

\subsection{ITA-Based Skin Type Misclassification}\label{sec:misclassification}

To convert the inter-device ICC into a clinical-deployment metric, we mapped each subject-region's post-CCM ITA to the six-category skin type classification of Del Bino and Bernerd\cite{DelBino2013} and compared the device-assigned category against the DSLR reference. Six-category boundaries in ITA are $-30$\textdegree, 10\textdegree, 28\textdegree, 41\textdegree, and 55\textdegree, separating Dark, Brown, Tan, Intermediate, Light, and Very light in ascending order, with boundary values assigned to the lower-ITA bin. Unlike ITA, Melanin Index lacks a consensus-based categorical scheme. To avoid introducing arbitrary cutpoints, its inter-device reliability is evaluated exclusively on a continuous scale via ICC, as detailed in Table~\ref{tab:clinical_icc}. Aggregate misclassification rates using global CCM and region-specific CCM are reported in Table~\ref{tab:misclassification}. Overall misclassification rates using global CCM are \misclassTabletGlobal\% (tablet) and \misclassSmartphoneGlobal\% (smartphone). The high baseline rate reflects this cohort's mean post-CCM ITA sitting near the Tan/Intermediate boundary, where modest residuals shift many subjects across an adjacent category boundary even though the underlying ICC sits in the good band (Table~\ref{tab:clinical_icc}). ICC quantifies rank-order preservation across devices, whereas misclassification quantifies absolute category assignment relative to a fixed cutpoint. The gap between the two metrics reflects their different measurement targets rather than an inconsistency in the data. Per-region rates using global CCM are roughly uniform (\misclassPerRegionLowGlobal--\misclassPerRegionHighGlobal\%), and nearly all misclassifications (\misclassAdjacentPct\%) involve adjacent ITA categories such as ``Tan'' $\leftrightarrow$ ``Intermediate'', indicating systematic boundary shifts rather than gross errors. Region-specific calibration cuts the overall rate to \misclassTabletRegional\% (tablet) and \misclassSmartphoneRegional\% (smartphone), with the chin region reaching the lowest regional-CCM rate among all five regions (\misclassChinTabletRegional\% for both consumer devices). The forehead's smartphone misclassification rate drops only marginally using regional calibration (\misclassForeheadSmartphoneGlobal\% to \misclassForeheadSmartphoneRegional\%), whereas chin and glabella drop from \misclassChinSmartphoneGlobal\% and \misclassGlabellaSmartphoneGlobal\% to \misclassChinSmartphoneRegional\% and \misclassGlabellaSmartphoneRegional\%, respectively.

\begin{table}[ht]
\caption{Per-region ITA-based skin type misclassification rates using global versus region-specific Color Correction Matrix (CCM) normalization (n=\nFrontPerDevice samples per device pair; n=965 per region). Classifications follow the Del Bino and Bernerd\cite{DelBino2013} six-category ITA scheme. The final row reports the overall pooled rate across all five regions.}\label{tab:misclassification}
\centering
\begin{tabular}{lcccc}
\toprule
 & \multicolumn{2}{c}{\textbf{Tablet vs DSLR}} & \multicolumn{2}{c}{\textbf{Smartphone vs DSLR}} \\
\cmidrule(lr){2-3} \cmidrule(lr){4-5}
\textbf{Region} & Global (\%) & Regional (\%) & Global (\%) & Regional (\%) \\
\midrule
Forehead    & \misclassForeheadTabletGlobal   & \misclassForeheadTabletRegional   & \misclassForeheadSmartphoneGlobal   & \misclassForeheadSmartphoneRegional \\
Left cheek  & \misclassLcheekTabletGlobal     & \misclassLcheekTabletRegional     & \misclassLcheekSmartphoneGlobal     & \misclassLcheekSmartphoneRegional \\
Right cheek & \misclassRcheekTabletGlobal     & \misclassRcheekTabletRegional     & \misclassRcheekSmartphoneGlobal     & \misclassRcheekSmartphoneRegional \\
Chin        & \misclassChinTabletGlobal       & \misclassChinTabletRegional       & \misclassChinSmartphoneGlobal       & \misclassChinSmartphoneRegional \\
Glabella    & \misclassGlabellaTabletGlobal   & \misclassGlabellaTabletRegional   & \misclassGlabellaSmartphoneGlobal   & \misclassGlabellaSmartphoneRegional \\
\midrule
Overall     & \misclassTabletGlobal           & \misclassTabletRegional           & \misclassSmartphoneGlobal           & \misclassSmartphoneRegional \\
\bottomrule
\end{tabular}
\end{table}

\section{Discussion}\label{sec:discussion}

\subsection{Achievable Inter-Device Reliability Under Standard Calibration}

The benchmark places consumer-camera reliability at a concrete operating point. Standard global CCM using canonical IEC 61966-2-1 sRGB delivers good inter-device agreement on both Melanin Index and ITA, with both indices clearing the 0.75 threshold widely used to denote clinical reliability.\cite{Portney2015} The remaining ITA shortfall using global CCM is dominated by anatomical region rather than by the source device (Section~\ref{sec:variance}). The next gain is thus available within the same hardware via region-aware calibration rather than through new imaging technology. Region-specific CCM raises ITA into the excellent band (Section~\ref{sec:regional}) and reduces clinical-decision misclassification at the same time (Section~\ref{sec:misclassification}), identifying region-aware calibration as the largest remaining source of improvement.

\subsection{Implications for Tele-Dermatology Deployment}

Three protocol implications follow. First, raw uncalibrated consumer-device measurements are clinically unusable in this benchmark, with no consumer-device-versus-DSLR pair achieving the $\Delta E < \deltaEThreshold$ threshold. Linear CCM normalization is therefore not optional. Second, the canonical IEC 61966-2-1 sRGB transfer function should be applied at decode time using a standard library implementation rather than a per-device gamma estimate, because the inter-device ICC values reported here all assume canonical sRGB inputs.\cite{IEC61966} Third, region-specific CCM trained per anatomical region delivers the largest single improvement on top of standard linear calibration. Coupling this with facial-landmark detection enables automatic per-region correction that requires no additional hardware beyond what consumer cameras already provide.\cite{Song2019,Chopra2015,Pan2020}

\subsection{Limitations}\label{sec:limitations}

The benchmark has scope limitations that affect generalizability. Non-linear and biomarker-aware corrections (for example, polynomial or neural-network mappings, or CCM losses with explicit per-channel weighting) were not evaluated, because the primary objective was to characterize what standard linear pipelines deliver using canonical-sRGB inputs and the additional gain available from region-aware calibration. Such non-linear corrections would require larger training sets to avoid overfitting, particularly within the narrow $L^*$ range of this cohort. Five-fold cross-validation assessed within-dataset generalization, but external validation on new subjects or different lighting conditions was not performed. By construction, the clinical indices analyzed here are deterministic CIELAB formulas. ITA is the angular skin-type correlate adopted in dermatology,\cite{Chardon1991,DelBino2006} and the Melanin Index is the standard CIELAB surrogate for skin pigmentation following Takiwaki\cite{Takiwaki1998} and its digital-imaging extension by Yamamoto et al.\cite{Yamamoto2008} The inter-device reliability values reported here therefore quantify how reliably these standard formulas transfer across consumer hardware rather than absolute biomarker validity, which would require a ground-truth reference such as reflectance spectroscopy or histology and is outside the scope of this benchmark.

The cohort is limited to Korean subjects (mean reference $L^* \approx \meanLstarRef$ using canonical sRGB), restricting direct generalization to lighter- and darker-skinned populations. The dataset is restricted to healthy skin, and pathological tissue may exhibit different spectral properties through altered melanin density, vascular proliferation, or stratum corneum changes that differentially affect specific CIELAB channels.

Regarding experimental conditions, the AI Hub dataset documents the DSLR acquisition room as containing unspecified lighting equipment and the tablet and smartphone acquisition area as having generic ceiling lighting in a waiting room. Specific illuminant spectrum, color temperature, and illuminance are not disclosed.\cite{AIHub2023} The consumer-device images therefore reflect routine indoor lighting rather than a calibrated optical rig, strengthening external validity for clinic deployment while limiting the ability to disentangle hardware-driven from illumination-driven variance. The DSLR reference is not a spectrophotometric ground truth, and its colorimetric error ($\Delta E \approx 1$--$3$) sets a floor on ICC precision. The DSLR multi-angle self-agreement (ICC = \iccDslrCeilingLow--\iccDslrCeilingHigh, Supplementary Fig.~\ref{fig:dslr_ceiling}) is computed across the seven viewing angles of the DSLR acquisition protocol and therefore absorbs within-subject angular variance, whereas the cross-device ICCs in Table~\ref{tab:clinical_icc} use front-facing captures only. The multi-angle ceiling bounds the reference's own noise budget under angular variation and exceeds the raw and global-CCM cross-device ICCs. Regional CCM (MI ICC = \iccMelaninRegional, ITA ICC = \iccItaRegional) crosses the ceiling because per-region calibration removes the regional variance component that the multi-angle ceiling absorbs. A camera-based reference was preferred over a contact spectrophotometer because it matches the optical quantity measured by the consumer-device pipeline under evaluation.

The regional variance pattern reported in Section~\ref{sec:variance} (chin highest, forehead lowest in within-subject angular variability, Supplementary Fig.~\ref{fig:region_angular}) is consistent with the geometric sensitivity of curved facial surfaces to small viewing-angle shifts, but the dataset does not separately control for region-specific differences in capture motion, illumination geometry, or skin specularity. The dominant regional mechanism is therefore not identified. The practical conclusion that region-aware calibration recovers excellent inter-device reliability (Section~\ref{sec:regional}) does not depend on resolving this.

\section*{Data Availability}

The Korean Skin Condition Measurement dataset used in this study is available through AI Hub (\url{https://aihub.or.kr}) under restricted access. The minimal dataset required to replicate the findings comprises matched facial images from DSLR, tablet, and smartphone devices with associated subject metadata (age group, sex, skin condition class) and facial region annotations. Access requires registration and approval through AI Hub, subject to a data use agreement. Processed summary statistics and analysis outputs are available in the Supplementary Information and the code repository (\url{https://github.com/hpicsk/regional-ccm}).

\section*{Code Availability}

The underlying code for this study is available at \url{https://github.com/hpicsk/regional-ccm} (MIT License). The repository includes scripts for canonical-sRGB color extraction, global and region-specific Color Correction Matrix normalization, clinical index calculation, ICC reliability analysis, and analysis of variance.

\section*{Supplementary Information}

Supplementary figures, tables, and an appendix reproducing relevant excerpts of the AI Hub usage guideline are provided as an appendix following the references.

\section*{Funding}

This study received no funding.

\section*{Author Contributions}

S.K. conceived the study, designed the methodology, performed the analysis, and wrote the manuscript. J.-K.K. reviewed the manuscript and provided critical feedback. All authors read and approved the final manuscript.

\section*{Competing Interests}

The authors declare no financial or non-financial competing interests.

\section*{Ethics Approval}

This study is a secondary analysis of the publicly available Korean Skin Condition Measurement dataset obtained from AI Hub (aihub.or.kr). The original data acquisition was conducted under an Institutional Review Board (IRB)--approved protocol in accordance with the Korean Bioethics and Safety Act, as documented in the dataset's usage guideline (v3.5, \S5.2). The present analysis falls outside the statutory definition of ``human research'' under Article 2(2) of the Enforcement Decree of the Korean Bioethics and Safety Act because it used only pre-existing, publicly released, de-identified data and involved no new contact with human participants, and was therefore exempt from additional IRB review.

\section*{Consent to Participate}

Informed consent was obtained from all participants by the original data collectors as part of the AI Hub data collection protocol.

\section*{Use of Large Language Models}

A large language model (Claude Opus 4.6, Anthropic) was used for the following tasks: (1) drafting manuscript prose from author-provided outlines and structured notes, (2) editing for clarity and concision, and (3) writing Python analysis scripts from author-specified requirements and pseudocode. All LLM-generated content was reviewed, verified, and approved by the authors. The LLM did not contribute to study design, data collection, or interpretation of results.

\bibliographystyle{srt-ama}
\bibliography{references}

\clearpage


\appendix
\renewcommand{\thesection}{\Alph{section}}
\renewcommand{\thesubsection}{\thesection.\arabic{subsection}}
\setcounter{section}{0}

\renewcommand{\thefigure}{S\arabic{figure}}
\renewcommand{\thetable}{S\arabic{table}}
\renewcommand{\theHfigure}{S\arabic{figure}}
\renewcommand{\theHtable}{S\arabic{table}}
\setcounter{figure}{0}
\setcounter{table}{0}

\section*{Supplementary Figures}

\begin{figure}[htbp]
\centering
\includegraphics[width=0.9\textwidth]{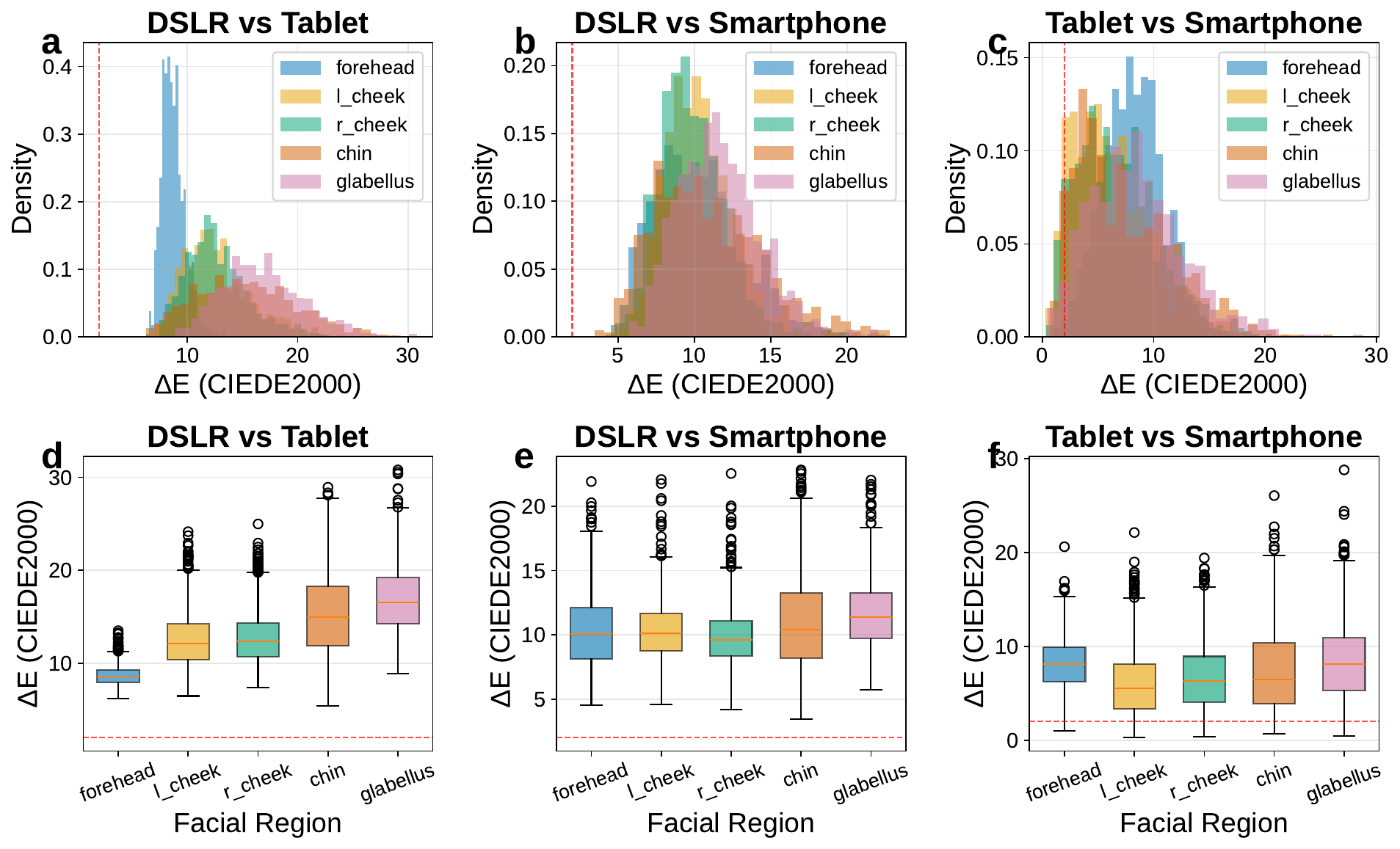}
\caption{\textbf{CIEDE2000 color difference ($\Delta E$) distributions by facial region.} Overlaid density histograms (top row) and box plots (bottom row) of pre-normalization $\Delta E$ between consumer devices and digital single-lens reflex (DSLR) reference, stratified by facial region (forehead, blue; left cheek, orange; right cheek, green; chin, vermillion; glabella, pink). Box plots show median (horizontal line), interquartile range (IQR; box), whiskers extending to 1.5 times IQR, and outliers (circles). n=\numSubjects matched pairs per region per device pair. Red dashed vertical line (top) and red dashed horizontal line (bottom) indicate the clinical acceptability threshold ($\Delta E = 2$). The glabella and chin regions exhibit the largest color differences in the tablet-vs-DSLR comparison, while smartphone-vs-DSLR differences are similar across regions. The analysis of variance (ANOVA) identifies anatomical region as the dominant source of variance in pre-calibration disagreement.}
\label{fig:dE_by_region}
\end{figure}

\begin{figure}[htbp]
\centering
\includegraphics[width=0.9\textwidth]{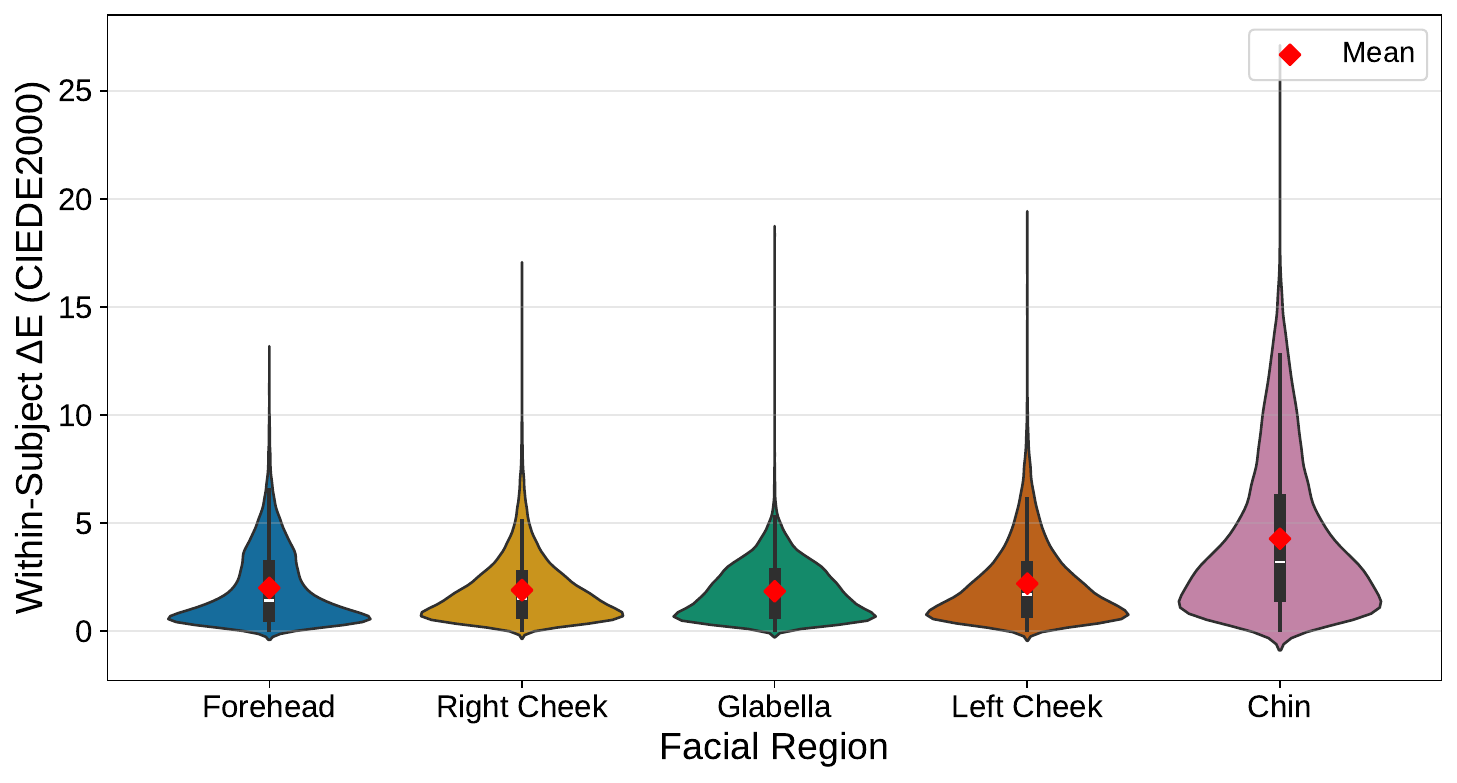}
\caption{\textbf{Within-subject color variability across digital single-lens reflex (DSLR) capture angles by facial region.} Violin plots (color-coded by region in ascending median order: forehead, blue; right cheek, gold; glabella, green; left cheek, vermilion; chin, pink; kernel density estimation) with overlaid box plots showing the distribution of pairwise $\Delta E$ (CIEDE2000) computed across 7 viewing angles for the same subject-region pair (n=\numSubjects subjects, full cohort). The chin region exhibits substantially higher within-subject angular variability than the forehead. Red diamonds indicate mean values. SD, standard deviation.}
\label{fig:region_angular}
\end{figure}

\begin{figure}[htbp]
\centering
\includegraphics[width=0.9\textwidth]{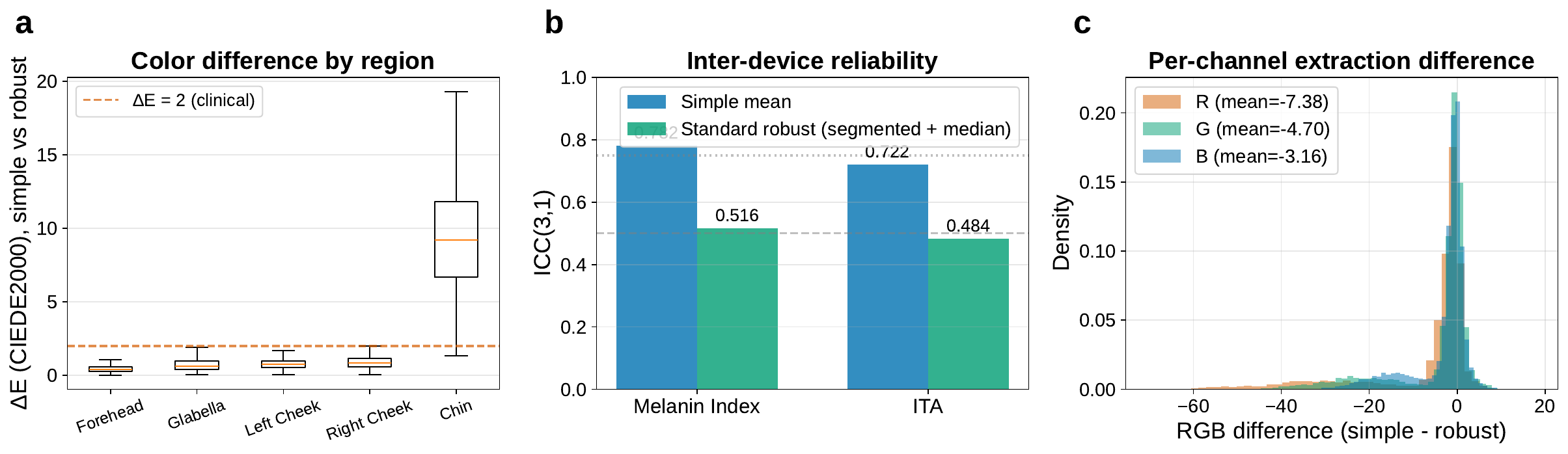}
\caption{\textbf{Robust skin-color extraction alters the chin region while leaving other facial regions unchanged.} Sensitivity analysis comparing simple full-ROI mean extraction (primary method) against a robust pipeline combining YCbCr skin-pixel segmentation,\cite{ChaiNgan1999} dichromatic specular/diffuse separation,\cite{Shafer1985,Bahmani2021} and a per-channel median (n=\nCrossDeviceTotal region-of-interest [ROI] extractions). a) CIEDE2000 color difference between the two extractions by facial region. The forehead, glabella, and cheeks agree (mean $\Delta E \leq \meanDeltaERoiNonChin$) while the chin diverges (mean $\Delta E = \meanDeltaERoiChin$). b) Intraclass correlation coefficient (ICC(3,1)) for Melanin Index (MI) and Individual Typology Angle (ITA) under each extraction. Pooled ICC declines from \iccRobustRoiMiSimple\ to \iccRobustRoiMiRobust\ (MI) and \iccRobustRoiItaSimple\ to \iccRobustRoiItaRobust\ (ITA), driven by reduced between-subject variance at the chin rather than increased cross-device disagreement. c) Per-channel RGB differences between extraction methods. DSLR, digital single-lens reflex; RGB, red--green--blue.}
\label{fig:robust_roi}
\end{figure}

\begin{figure}[htbp]
\centering
\includegraphics[width=0.9\textwidth]{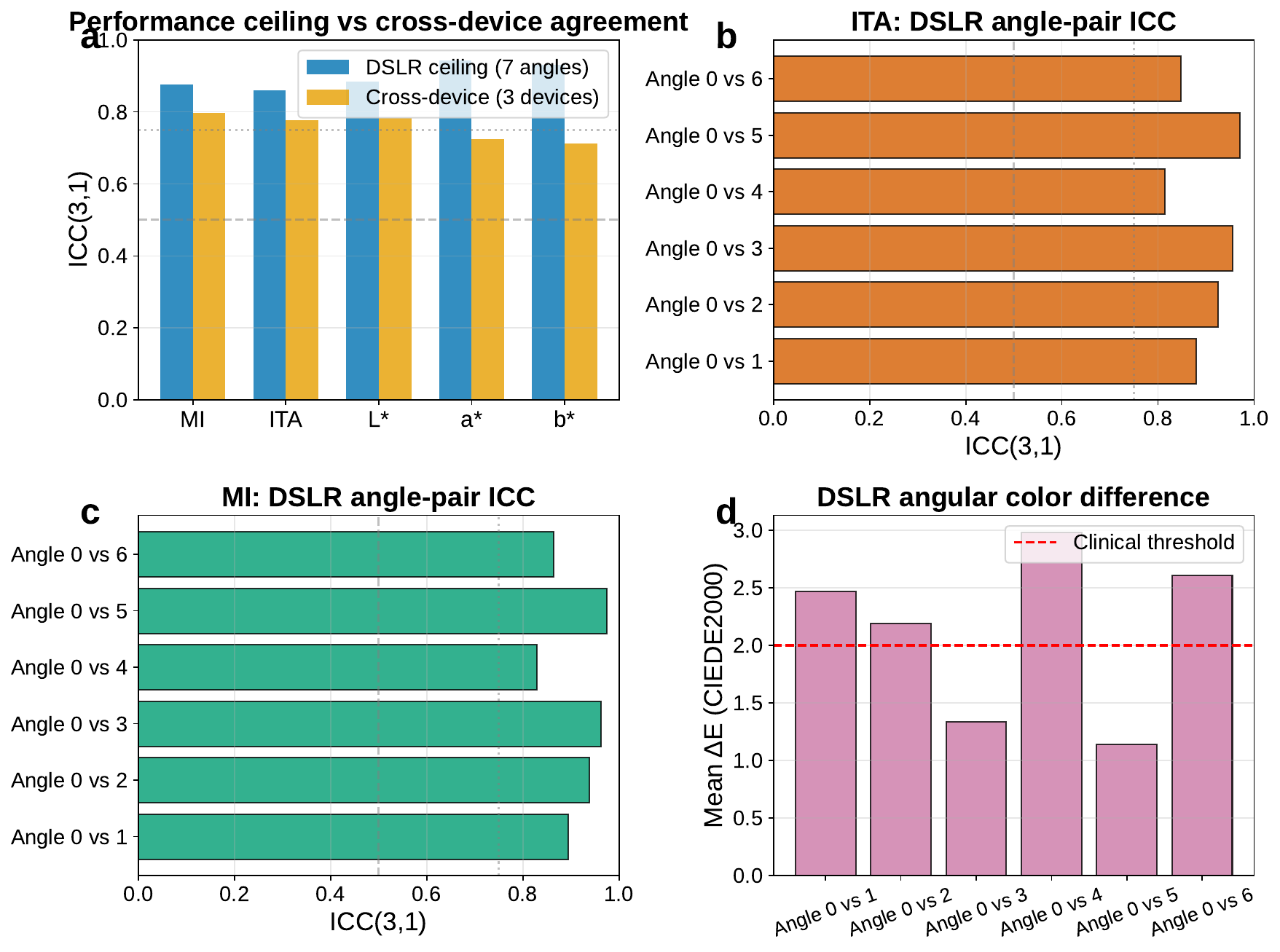}
\caption{\textbf{Digital single-lens reflex (DSLR) multi-angle self-agreement establishes a high performance ceiling (ICC 0.86--0.94), confirming that cross-device disagreement reflects device differences rather than reference noise.} a) Comparison of DSLR ceiling ICC(3,1) (\numDslrAngles viewing angles as ``raters''; blue bars) against cross-device ICC (3 devices, before CCM normalization; orange bars) for clinical indices and Commission Internationale de l'\'Eclairage L*a*b* (CIELAB) channels (n=\nFrontPerDevice per device). The DSLR ceiling exceeds raw cross-device agreement for all indices. b) Pairwise ITA ICC for DSLR angle 0 versus each off-angle capture. c) Pairwise Melanin Index (MI) ICC by angle pair. d) Mean DSLR angular $\Delta E$ (CIEDE2000) per angle pair (mean across all pairs $\approx \meanDslrAngularDeltaE$). Per-pair values exceed the clinical acceptability threshold ($\Delta E = 2$, red dashed line) for the wider rotations (\viewAngleSmall--\viewAngleLarge\textdegree{} off-frontal), consistent with the analysis-of-variance finding that anatomical region and viewing geometry are dominant pre-CCM variance contributors. The DSLR ceiling ICCs in panels a--c remain in the good-to-excellent range, indicating that even at this within-device $\Delta E$ level the rank-order of subjects is preserved across viewing angles. CCM, Color Correction Matrix; ICC, intraclass correlation coefficient; ITA, Individual Typology Angle.}
\label{fig:dslr_ceiling}
\end{figure}

\end{document}